\documentclass[12pt]{article}
\usepackage{a4}
\usepackage{rotate}
\usepackage{epsfig}
\usepackage{amsmath}
\usepackage{cite}

\newcommand{\de}{\partial}
\newcommand{\lsi}{\raise0.3ex\hbox{$<$\kern-0.75em\raise-1.1ex\hbox{$\sim$}}}
\newcommand{\gsi}{\raise0.3ex\hbox{$>$\kern-0.75em\raise-1.1ex\hbox{$\sim$}}}
\newcommand{\lsim}{\mathop{\lsi}}

\makeatletter \@addtoreset{equation}{section} \makeatother

\begin{document}
\setlength{\baselineskip}{0.6cm}
\begin{titlepage}
\begin{flushright}
HD-THEP-01-001\\
IEM-FT-209/01\\
IFT-UAM/CSIC-01-01\\
hep-ph/0101249\\         
\end{flushright}
\begin{centering}
\vfill
{\bf Bubble Walls, CP Violation and Electroweak Baryogenesis in the MSSM}
\vspace{.5cm} \\
S.J. Huber$^{a,}$\footnote{shuber@bartol.udel.edu}
P. John$^{b,}$\footnote{john@makoki.iem.csic.es} and
M.G. Schmidt$^{c,}$\footnote{m.g.schmidt@thphys.uni-heidelberg.de}
\vspace{.5cm} \\ {\em
$^a$Bartol Research Institute, 217 Sharp Lab, Newark, DE 19716, USA\\
$^b$Instituto de Estructura de la Materia (CSIC), Serrano 123, 28006
Madrid, Spain\\
$^c$Institut f\"ur Theoretische Physik,
Philosophenweg 16,
D-69120 Heidelberg, Germany
}

\vspace{2cm}

{\bf Abstract}

\vspace{0.5cm}

\end{centering}

We discuss the generation of the baryon asymmetry by a strong first
order electroweak phase transition in the early universe, particularly
in the context of the MSSM. This requires a thorough numerical
treatment of the bubble wall profile in the case of two Higgs
fields. CP violating complex particle masses varying with the Higgs
field in the wall are essential. Since in the MSSM there is no
indication of spontaneous CP violation around the critical temperature
(contrary to the NMSSM) we have to rely on standard explicit CP
violation. Using the WKB appro\-xi\-mation for particles in the plasma
we are led to Boltzmann transport equations for the difference of
left-handed particles and their CP conjugates.  This asymmetry is
finally transformed into a baryon asymmetry by out of equilibrium
sphaleron transitions in the symmetric phase. We solve the transport
equations and find a baryon asymmetry depending mostly on the CP
violating phases and the wall velocity.

\vspace{0.3cm}\noindent

\vfill \vfill
\noindent

\end{titlepage}
\newpage
\section{Introduction}
\label{intro}
The electroweak interactions violate baryon number. There is also a
source of CP violation in the Standard Model (SM), and possibly even
stronger ones in models beyond the SM, like supersymmetric theories.
Thus it is very tempting to expect that these theories which are
presently under stringent experimental tests allow to explain the
creation of a baryon asymmetry in the early universe. However, 
in thermal equilibrium a baryon asymmetry ($B+L\neq 0$) is
reduced by the electroweak interactions (conserving $B-L$). If there
are no $B-L$ violating interactions at the GUT scale creating $B-L\neq
0$, (as e.g. in the recently discussed ``leptogenesis'' models
\cite{Leptogen}) one needs a strong deviation from equilibrium to
create a baryon asymmetry at the electroweak scale and fast freeze out
in the subsequent quasi equilibrium. Thus a strong first order phase
transition with bubble nucleation is needed.  However, combined
analytical and lattice based work in the last years has made sure that
such a phase transition is not present in the SM; actually there is no
phase transition at all for Higgs masses beyond about the $W$-boson
mass\cite{klrs}.  Thus, one has to inspect variants of the
SM. Supersymmetric models are the most attractive ones. In particular
the Minimal Supersymmetric Extension (MSSM) closest to the SM is
investigated most intensively in experiments. With a light stop
superpartner of the right handed top having a mass some GeV below the
top mass, one can increase the $\phi^3$-term in the effective thermal
Higgs potential leading to a strong first order phase
transition\cite{2loop,CarlosEspinosa,litestop,Cline12,MooreServant}.

In order to create a baryon asymmetry we first want  to generate a
chiral asymmetry in the hot phase in front of the bubble which is then
transformed into a baryon asymmetry through hot electroweak
sphalerons. For this we need CP violation in the Higgs field bubble
wall or some CP violation in the hot plasma disturbed by the wall.
As consequence, sources are induced in the transport equations. 
In the MSSM the chargino
and the stop mass matrices contain explicitly CP violating phases. The
Higgs Lagrangian can obtain CP violating loop effects but we will also
discuss in some detail the attractive possibility of a spontaneously
generated ``transitional'' CP violation just in the bubble wall around
the critical temperature
\cite{MaPo,fkot,Mikko3dCP,QuirosCP,CKVischer,HuJoSchmiLai,LaiRuCPlat}.
Therefore general methods for the calculation of multi-scalar bubble
walls are necessary. In Section~\ref{sec:eqs} we derive the
relevant equations of motion, including CP violation, and present
in Section~\ref{sec:howto} a method to solve them in general. In
Section~\ref{sec:CPgeneral} we present their applications to the
considered supersymmetric models.

Further on, in Section~\ref{sec:BAU} we will study the transport
phenomena close to the bubble wall in the quasi classical particle
``WKB''-approximation\cite{CKN,CJK,JPT1,CJKneu}.  Thus we neglect soft
boson classical field effects, which for example may play a role for
the friction due a gauge field in the hot plasma \cite{GMoo}, as well
as off-shell memory effects to be discussed in the framework of
Quantum-Boltzmann equations
\cite{QBE,irgendwas_von_Riotto,non-eq,seco00}.  The stationary
velocity of the bubble wall is very important for the finally created
baryon asymmetry. It is related to the friction term in the transport
equations. This has been treated in ref.~\cite{MP12} and recently in
\cite{JSwall} pointing to a very small velocity in the MSSM which, in
general, is quite effective for producing an asymmetry.  It turns out
that it might even be too small for a maximum exploit\cite{ariel}.  In
ref.~\cite{HS} a class of supersymmetric models with an additional
gauge singlet has been worked out. These allow for a strong first
order phase transition already at tree level. In our discussing of
wall profiles and CP violation we will also include this interesting
case.  Concerning the transport equations in the WKB approximation we
can return from the calculation in ref.~\cite{HS00} including the
singlet field to a critical discussion of the MSSM case which was
already worked out in ref.~\cite{CJK} before and reconsidered recently
\cite{CJKneu}. Finally we give numerical results for the baryon
asymmetry in the given framework.
%
%
%
%
\section{Field Equations and CP Violation}
\label{sec:eqs}
For our purposes it is sufficient to consider the one-loop corrected
effective Higgs potential $V_T$ with thermal masses. We use the standard
notation for the MSSM \cite{HK}.
In general, there are CP violating phases between the Higgs field moduli  
$h_1$ and  $h_2$.
Due to gauge invariance, the effective Higgs potential only
depends on one combination $\theta=\theta_1+\theta_2$ of the phases,
\begin{equation}
V_T(H_1,H_2)=V_T(h_1,h_2,\theta).
\end{equation}
Defining
\begin{equation} \label{thetadef}
\bar\theta=\theta_1-\theta_2,
\end{equation}
we can rewrite the kinetic terms for the Higgs bosons in the
Lagrangian density
\begin{equation} \label{thlag}
{\cal  
L}_{kin}=\frac{1}{2}\partial_{\mu}h_1\partial^{\mu}h_1+\frac{1}{2}\partial_{\mu}h_2\partial^{\mu}h_2
+\frac{h_1^2+h_2^2}{8}(\partial_{\mu}\theta\partial^{\mu}\theta+
\partial_{\mu}\bar\theta\partial^{\mu}\bar\theta)
+\frac{h_1^2-h_2^2}{4}\partial_{\mu}\theta\partial^{\mu}\bar\theta.
\end{equation}
Since $\partial_{\bar\theta}V_T=0$ the Euler-Lagrange equation
for $\bar\theta$ is
\begin{equation} \label{thconst}
 (h_1^2+h_2^2) \partial^{\mu}\bar\theta+ (h_1^2-h_2^2) \partial^{\mu}\theta=C_0
\end{equation}
which minimizes the energy for $C_0=0$.

After elimination of $\partial_{\mu}\bar\theta$
(\ref{thconst}), the equations of motion for $h_1$, $h_2$ and $\theta$
read
\begin{eqnarray}
\label{h1eq}
\partial_{\mu}\partial^{\mu}h_1+\frac{h_1h_2^4}{(h_1^2+h_2^2)^2}
\partial_{\mu}\theta\partial^{\mu}\theta+\frac{\partial}{\partial h_1}V_T=0,
\label{eqmo1}\\
\nonumber \\
\partial_{\mu}\partial^{\mu}h_2+\frac{h_2h_1^4}{(h_1^2+h_2^2)^2}
\partial_{\mu}\theta\partial^{\mu}\theta+\frac{\partial}{\partial h_2}V_T=0, 
\label{eqmo2}\\
\nonumber \\
\partial_{\mu}\left[\frac{h_1^2h_2^2}{h_1^2+h_2^2}\partial^{\mu}\theta\right]
+\frac{\partial}{\partial \theta}V_T=0.\label{eqmo3}
\end{eqnarray}

Two limits are important. The first limit corresponds to a
radially symmetric spatial solution of
\begin{equation}
\frac{\de^2h_i}{\de r^2} -\frac{2}{r}\frac{\de h_i}{\de r}-\frac{\de  
V_T(h_i)}{\de h_i}=0
\label{critbub}
\end{equation}
(without CP phase $\theta$).  The solution describes the initial state
of nucleating bubbles, the ``critical bubble''. Eqs.~(\ref{critbub})
are also to be solved for the determination of the tunneling
probability\cite{Linde,AndersonHall,LMT}.

Second, constraining equations (\ref{eqmo1})-(\ref{eqmo3}) to a
stationary wall (``domain wall'') moving with velocity $v_w$ we are
left with just one direction $x=z-v_wt$ at late time $t$ perpendicular
to the wall. This is the more important period for baryogenesis.
With the assumption of stationarity and almost planar bubble walls
the equations
(\ref{eqmo1})-(\ref{eqmo3}) reduce to
\begin{eqnarray}
h_1^{\prime\prime}+\frac{h_1h_2^4}{(h_1^2+h_2^2)^2}\theta^{\prime\prime}
-\frac{\partial}{\partial h_1}V_T=0,\label{eqmo1N}\\
h_2^{\prime\prime}+\frac{h_2h_1^4}{(h_1^2+h_2^2)^2}\theta^{\prime\prime}
-\frac{\partial}{\partial h_2}V_T=0,\label{eqmo2N}\\
\de_x\left[\frac{h_1^2h_2^2}{h_1^2+h_2^2}\theta'\right]
-\frac{\partial}{\partial \theta}V_T=0,\label{eqmo3N}
\end{eqnarray}
where the prime denotes $\de_x$. $V_T$ is deformed by the plasma (moving
in the wall frame) friction and has degenerate minima in the stationary
case like $V_{T=T_c}$. We roughly identify the deformed  $V_T$ with
$V_{T=T_c}$.

For the NMSSM singlet field $S$ it is more
 convenient \cite{HuJoSchmiLai, HS00} to divide it up
into real and imaginary components,
\begin{equation}
S=n+ic,
\end{equation}
implicitly introducing a phase variable
$\theta_S=\mathop{\arctan}(n/c)$.  The equations of motion for $n$ and
$c$ are of Klein-Gordon type
\begin{eqnarray} \label{seq}
\de_\mu\de^\mu n+\frac{\partial}{\partial n}V_T=0, \label{eqmo4}\\
\nonumber \\
\de_\mu\de^\mu c+\frac{\partial}{\partial c}V_T=0,\label{eqmo5}
 \label{ceq}
\end{eqnarray}
where $V_T$ is the NMSSM potential \cite{HS00}.

In Sec.~\ref{sec:NMSSMCP} we will see that especially the
variations of the CP violating quantities $\theta$ and $c$ in
the bubble wall play a very important role in the generation of the
baryon asymmetry. But first we will discuss more generally how these
equations may be solved.
%
%
%
%
\section{How To Find Bubble Wall Profiles}
\label{sec:howto}
The first order phase transition is mediated by expanding bubbles.
In order to compute the baryon asymmetry, we should follow the
history of bubbles from the moment of nucleation, until the time when the
broken phase fills the Universe. After nucleation, there is in general
a long period of stationary growth. Nucleation is characterized by
time dependent solutions to the full equations of motion in a
background of in general non-trivial profiles for temperature, velocity
and chemical potential. In the SM such solutions were investigated in
\cite{rth}. We will not consider the full problem here but focus
on the profiles of stationarily expanding almost planar bubble walls
at the critical temperature $T_c$ which interpolate between two minima
of the effective potential. This case is more interesting case for
baryogenesis. In case of friction with the background plasma the
corresponding temperature is the nucleation temperature
$T_n$ \cite{JSwall}.

For the bubble wall profiles in the MSSM without CP violation there
exist numerical approaches to solve the problem of critical bubbles
with two Higgs fields in \cite{SecoNum} and more general in
\cite{Moore261}. In \cite{CKVischer}, the CP profile has been
investigated in the background of a fixed Higgs profile.  We will now
discuss methods which enable us to find profiles in more general
scenarios as shown in Sections~\ref{sec:numana} and
\ref{ssec:explicitM} as well as in
\cite{HuJoSchmiLai,JSwall}.

The usual method for the SM case with only one Higgs field is to solve
the corresponding single equation of type (\ref{eqmo4})
numerically by ``turning around'' the effective potential $V_T$ and
dealing $x$ for a time $t$. Then the problem can be regarded as an
initial value problem for the inverted potential $-V_T$.  Usually an
``over\-shoot\-ing-under\-shoot\-ing'' procedure can be applied. This
works well since there is only one direction in field space. Moreover,
it can be implemented quite simply by the standard Runge-Kutta-method.

The  situation  is  completely different  once   there are additional
directions in  field  space.  Again  one  can  consider the  analogous
mechanical problem with the turned  around potential. Neglecting first
derivatives which act like friction on the potential, the initial value
problem is equivalent to a mass point rolling from the top of one hill
(first minimum) along the ridge such that it  comes to rest on the top
of the   second hill (second   minimum). Small changes  in the initial
conditions lead to  a completely different shape  of the solution.  In
general it  is   not possible  to  know  the  initial conditions  with
sufficient accuracy  to find the   desired solution. Hence  we have to
device another method.  Here we use the method  of minimization of the
functional of the squared equations of motion.

A stationary problem for big bubbles can be reduced to a problem in
one spatial coordinate $x=z-v_wt$ for a constant wall velocity $v_w$.
A constant velocity is caused by friction with the background plasma
which ensures that the minima are degenerate
\cite{JSwall,MP12,LMT,CarrKa,Linde}.

To solve Eqs.~(\ref{eqmo1N})-(\ref{eqmo3N}) one has to find field
configurations for which
\begin{equation}
S=\int_{-\infty}^{+\infty}dx \left[E_1^2(x)+E_2^2(x)+E_3^2(x)\right]=0.\label{SM}
\end{equation}
This is achieved by searching for the absolute minimum of $S$. Here
$E_1(x)$, $E_2(x)$ and $E_3(x)$ are the left hand sides of
Eqs.~(\ref{eqmo1N}), (\ref{eqmo2N}), and (\ref{eqmo3N}), respectively.
For the NMSSM the extension is straightforward
\begin{equation}
S=\int_{-\infty}^{+\infty}dx  
\left[E_1^2(x)+E_2^2(x)+E_3^2(x)+E_4^2(x)+E_5^2(x)\right],\label{SN}
\end{equation}
where $E_4(x)$ and $E_5(x)$ are the squares of stationary versions of
the l.h.s. of Eqs.~(\ref{eqmo4}) and (\ref{eqmo5}):
\begin{eqnarray}
E_4(x)=n^{\prime\prime}-\frac{\de V_T}{\de n},\\
E_5(x)=c^{\prime\prime}-\frac{\de V_T}{\de c}.\\
\nonumber
\end{eqnarray}
This minimization method has also been successfully used in
\cite{SecoNum} for the critical bubble.  The approach of
\cite{Kusenko} is in principle also a minimization procedure and
therefore the following considerations are also applicable. The method
of \cite{Moore261,MooreServant} deals directly with the saddle point.
Using the minimization method we have to solve a boundary value
problem instead of an initial value problem. Thus we have to use an
ansatz which fulfills the boundary conditions for each function for
which we want to find the time development.

Our procedure works in two steps. The first crucial step is to find an
ansatz which is as close as possible to the exact solution.

\paragraph{Step One: Ans\"atze}
In \cite{SecoNum,John3,HuJoSchmiLai,Cline12} it was found that a kink is a
good ansatz for the Higgs fields.  We therefore use for $N$ Higgs
fields
\begin{equation}
\phi_i^{kink}=\frac{v_i}{2}\left(1+\tanh(\frac{x}{L_i}+\hat{x}_i)\right),
\quad i=1\ldots N.\label{kink}
\end{equation}
The $L_i$ and $\hat{x_i}$ are determined by minimizing (\ref{SM}) or
(\ref{SN}) which are built up from these configurations in the Eqs. of
motion like (\ref{eqmo4})-(\ref{eqmo5}).  This first step is cheap in
terms of computer time since it is only a $N$-dimensional minimization
and can be performed very fast. Minimizing with respect to a few
parameters is a very successful first step since it reduces the value
of (\ref{SM}) already significantly compared to a general function
which only fulfills the boundary conditions. Hence the convergence is
considerably improved.

The shape of the tunneling trajectory depends strongly on the CP-odd Higgs
mass parameter $m_A$. Small values of $m_A$ give a larger mixing of
the CP even Higgs fields and a sharper bending curve (see \cite{John3}).

\begin{figure}[ht]
\vspace*{1cm}
\begin{picture}(0,400)
\put(-20,164){\epsfysize=8.2cm\rotate[r]{\epsffile{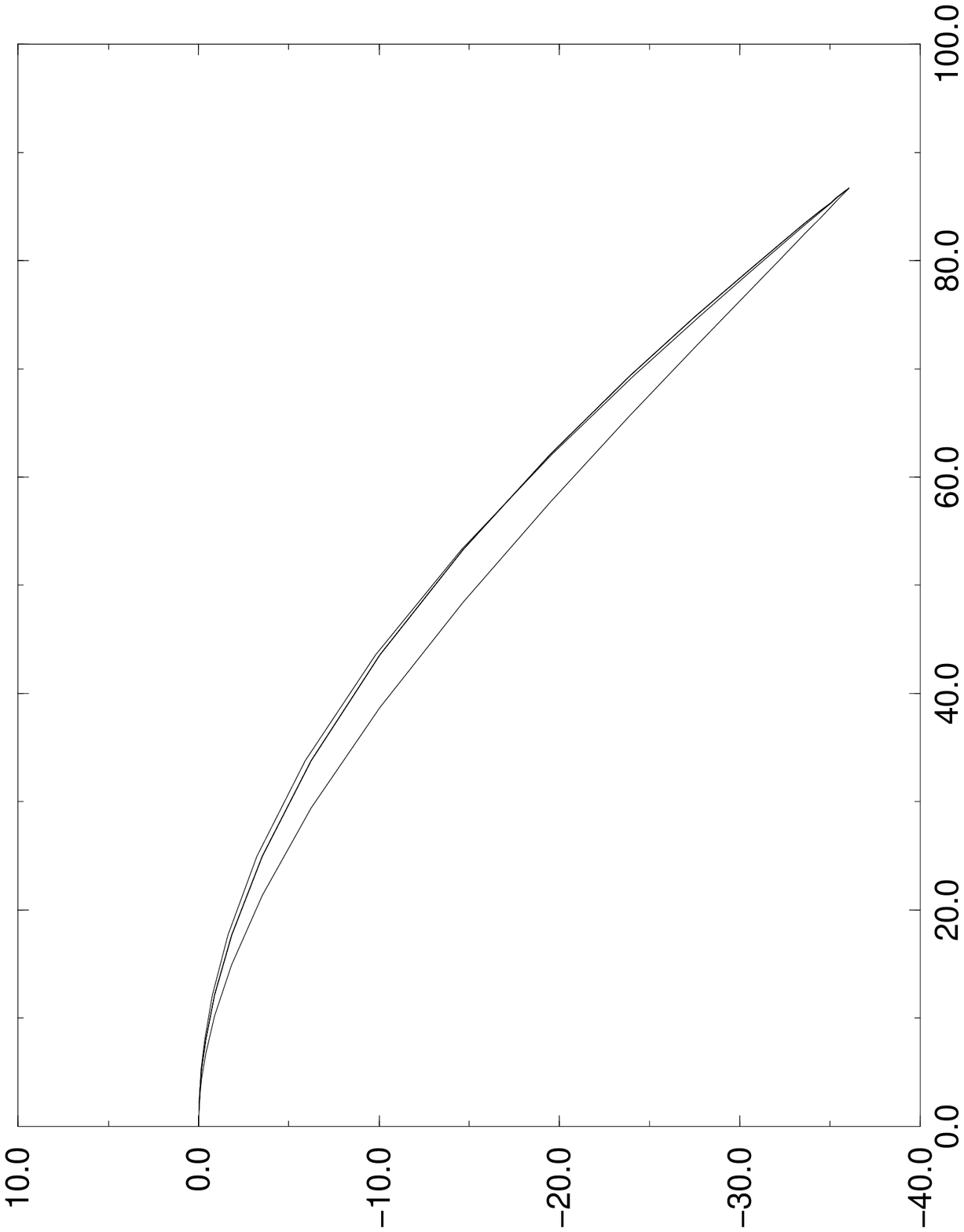}}}
\put(220,180){\epsfysize=7cm\rotate[r]{\epsffile{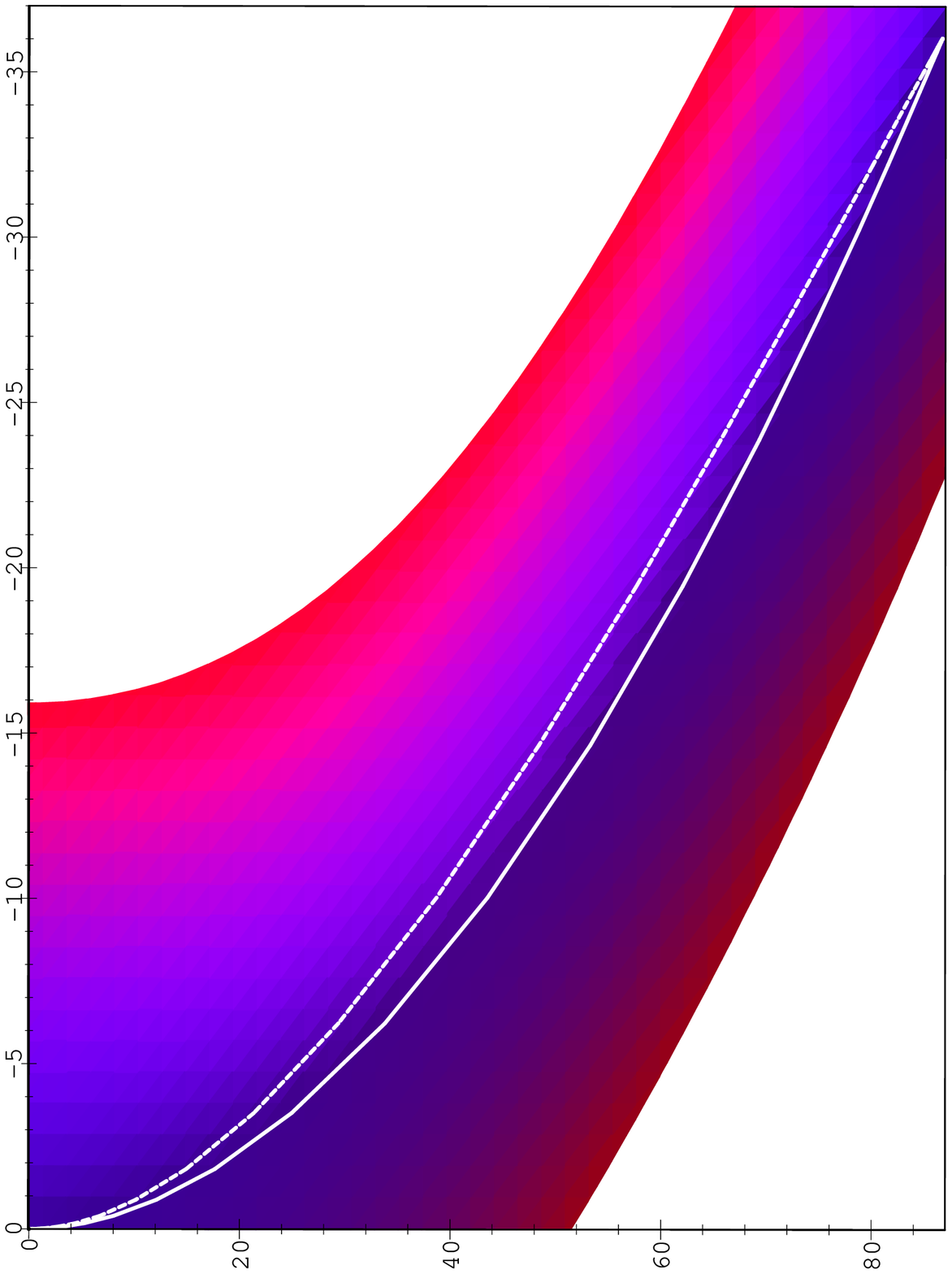}}}
\put(0,0){\epsfysize=6.8cm\rotate[r]{\epsffile{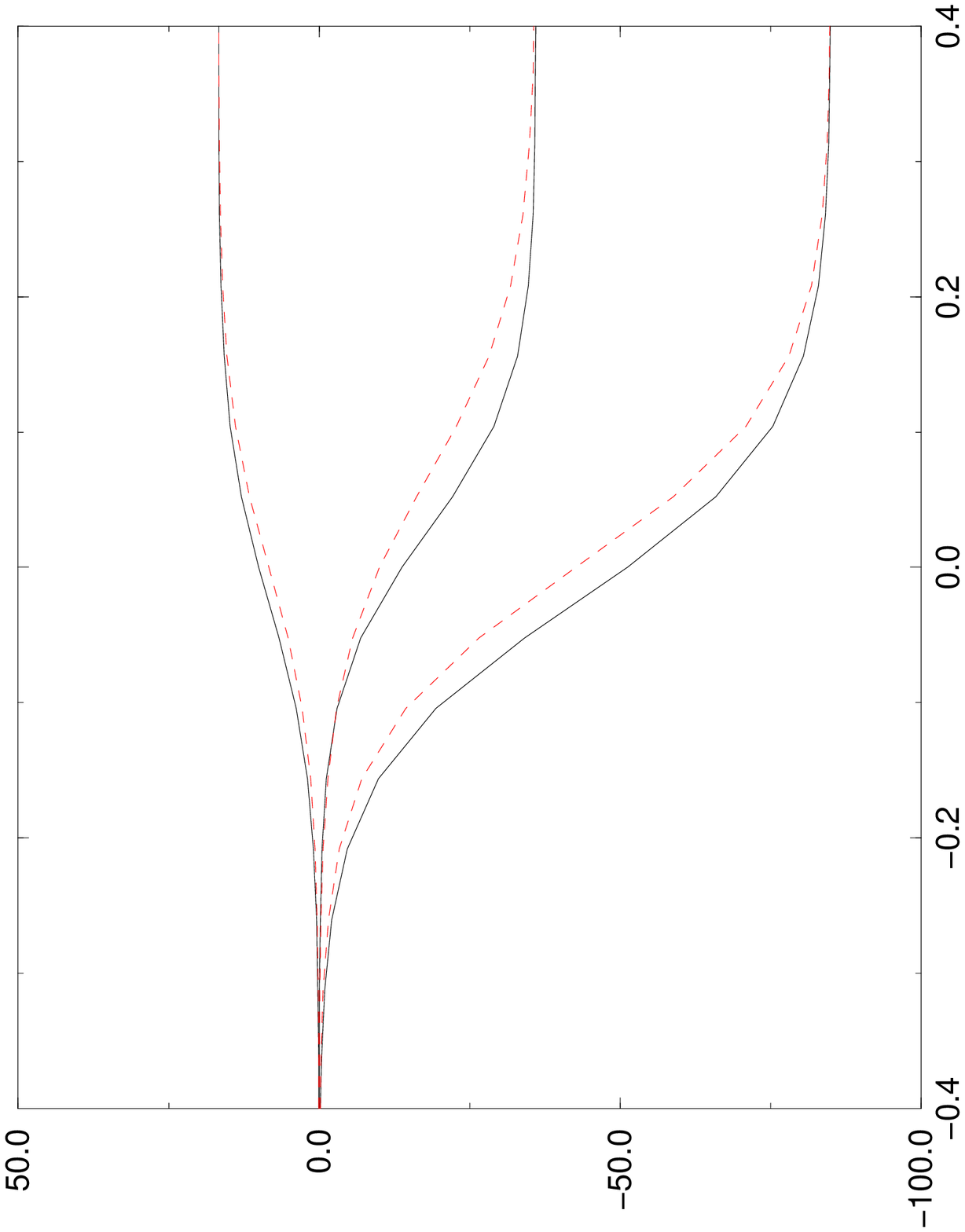}}}
\put(220,5){\epsfysize=7cm\rotate[r]{\epsffile{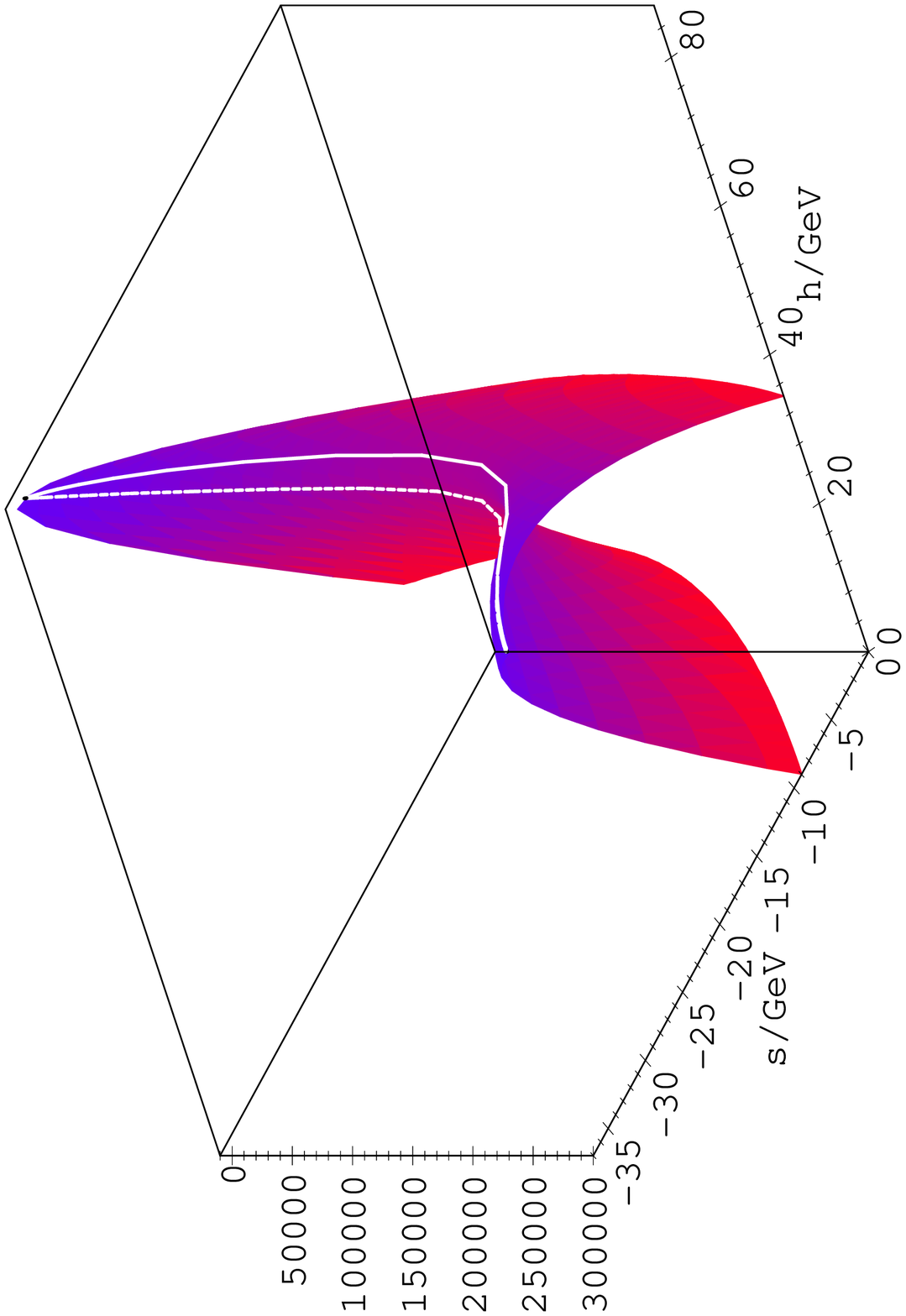}}}
\put(150,120){\tiny $h_1$}
\put(150,60){\tiny $h_2$}
\put(150,26){\tiny $s$}
\put(180,-5){\tiny $x/\rm{GeV}^{-1}$}
\put(0,10){\rotatebox{90}{\tiny $h_i(x)/$GeV}}
\end{picture}
\vspace*{1cm}
\caption{Solution and ans\"atze in the NMSSM. Left top: Fitted
$\tanh(x/L+{\hat x})$-ansatz and the ridge lying directly near the
actual solution. Lower left: Shape of solution versus $x$ compared to
a kink ansatz.  Right: 3-dimensional views of the effective potential with
solution and kink ansatz.}
\label{grat}
\begin{picture}(0,0)
\put(130,445){\tiny $h=\sqrt{h_1^2+h_2^2}/$GeV}
\put(-5,295){\rotatebox{90}{\tiny $s$/GeV}}
\put(400,440){\tiny $s/$GeV}
\put(210,295){\rotatebox{90}{\tiny $h/$GeV}}
\end{picture}
\end{figure}

Also the ridge between the the peaks turns out to be a rather fine
ansatz. In general, the ridge is difficult to define but often it can
be defined sufficiently in two scalar field dimensions. In a generic
NMSSM case the ridge is even nearer to the solution than a
superposition of two kinks. This is demonstrated in Fig.~\ref{grat},
where the solution and the ridge lie quite close to each other.

\paragraph{ridge} In two scalar dimensions an approximation to the 
ridge often can be defined as follows: Intersect the effective
potential in planes perpendicular to the straight connection between
the minima and take the maxima of the intersection curves in those
intersection planes as points of the ridge.\\
To use this in the NMSSM where we have basically 3 fields we define a
new average field $h=\sqrt{h_1^2+h_2^2}$ in the $h_1$-$h_2$ plane.
This is a good approximation for $h_1$ and $h_2$ which are close to a
straight line, similarly to the MSSM. They can subsequently be treated
as one. Then, a good approximation to the ridge can be found by
applying the above construction to the effective potential in the
$h-S$ space. In Sections~\ref{ssec:explicitM} we will discuss examples
which violate CP. In these cases the ridge cannot be defined in the
described simple manner. Fortunately these example cases again permit
ans\"atze utilizing the extended kink (\ref{kink}).

\paragraph{Step Two: Minimization and problems connected with it:}
With a preoptimized ansatz we come to the second step, the high
dimensional minimization. To represent (\ref{SM}) and (\ref{SN})
numerically we discretize them over a grid.  We have to
discretize the space variable $x=z-v_wt$, the fields $\phi_i(x_j)$,
the derivatives of the fields (first and second derivatives), and
derivatives of explicitly given functions like the effective
potential.  These points have recently been discussed partially in
\cite{John3,SecoNum,MooreServant}. We want to discuss here some
forthcoming details and computational problems.

The fields are defined on a grid of points $x_j$ along $x$. The first
derivatives turn out to be appropriately discretized by
$f'(x)=(f(x+\epsilon)-f(x-\epsilon))/\epsilon$ and the second
derivatives by
$f^{\prime\prime}(x)=(f(x+\epsilon)-2f(x)+f(x-\epsilon))/\epsilon^2$.
Accuracy is of order $\epsilon^2$ where $\epsilon$ is the grid
spacing. The discretized field equations (\ref{SM}) and (\ref{SN}) can
be built out of these elements and the discretized fields. Derivatives
including more points may even impair the result since they require
more additions and subtractions which are numerically problematic. For
this problem see also \cite{numrec,Knuth}. 

Minimization parameters are the values $\phi_i(x_j)$ and
$\theta(x_j)$, respectively. With $N$ fields defined at $M$ grid
points we have a $N\times M$-dimensional function to be minimized.

In our applications the minimization itself is accomplished using two
different methods, ``Powell's method'' \cite{numrec} and a
``Pseudo-Langevin'' method which will be discussed below.  We also
used the ``downhill simplex'' minimization method \cite{numrec} for
comparison with \cite{fkot} which turned out to be too slow for
practical use beyond a few computations. 

For the discussion of the practical usage of minimization we first
discuss a common problem of minimizations including numerical
derivatives.  In general several undesired minima occur which can be
categorized as follows:

a) Some are real solutions of the equations of motion. A trivial
example is $\theta(x)=0$, which always solves
(\ref{eqmo1})-(\ref{eqmo3}).

b) Fake minima due to the numerical representation of the
functional. This is a common problem of discretized derivatives with
finite differences.

c) Minimization of (\ref{SM}) is achieved by solving $\delta S=0$,
which is only a necessary condition for solution configurations. For a
squared form, $S=a^2$, also pseudo-solutions according to $\delta a=0$
may exist in addition to the (desired) solution $a=0$. Also combined
effects may occur as discussed in \cite{John3}.

Altogether this induces the problem that, starting from an ansatz
which is in the vicinity of such an apparent solution, the algorithm
might never converge to the desired exact solution. How can these
problems be solved?

One method is related to the method of ``Simulated Annealing''
\cite{numrec,Rothe}. There the trajectory through the high dimensional
configuration space is given by a Langevin type equation, e.g.
\begin{equation}
\frac{d q_i}{d\tau} = -\frac{\de S[q]}{\de q_i}+\eta_i(\tau),
\label{langevin}
\end{equation}
where the $q_i$ are the degrees of freedom of the discretized action
$S[q]$ (Eq.~(\ref{SM}) or (\ref{SN}))and $\eta_i(\tau_n)$ are random
variables, typically Gaussian distributed.  Every configuration is
randomly varied to jump out of unwanted local minima. For a global
progress in minimizing $S$ a small suffer loss is permitted. An
algorithmic temperature is defined which regulates the amplitudes of
the fluctuations.  This method is commonly used in lattice
calculations (e.g. \cite{GarciaBaal}).  Unfortunately, the action $S$
in our case is quite sensitive to randomly varying degrees of freedom
through the derivatives of the configurations. A random change in the
fields causes a strong change in the derivatives which increases
$S$. The convergence is very poor and it turns out that this method
does not solve our problem.

\begin{figure}[h]
\vspace*{-2.cm}
\hspace*{-2.5cm}
\begin{picture}(0,200)
\put(170,-50){\epsfysize= 7cm{\epsffile{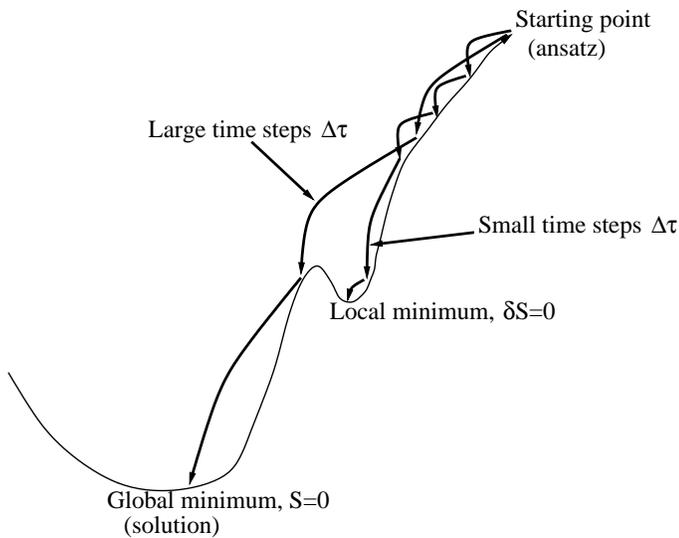}}}
\end{picture}
\vspace*{1.5cm}
\caption{Pseudo-Langevin method: Large time steps allow to overcome
small bumps and troughs. Small time steps or simple minimum finding
routines may lead into undesired local minima.}
\label{fig:bump}
\end{figure}

Writing down the discretization of (\ref{langevin}) and omitting
the fluctuation term we find the ``Pseudo-Langevin'' equation
\begin{equation}
q_i(\tau_{n+1})=q_i(\tau_n)-\Delta \tau \frac{\de S[q]}{\de q_i(\tau_n)}.
\label{PL}
\end{equation}
Now the new time interval $\Delta\tau$ of the new Hamiltonian system
is a common integration step for all degrees of freedom. The motion
promoted by Eq.~(\ref{PL}) is a motion without inertia through the
configuration space. It converges towards a minimum of $S$.  $\Delta
\tau$ is determined through the scale of $\de S[q]/\de q_i(n)$:
\begin{equation}
\Delta\tau \sim \frac{\de S[q]}{\de q_i(n)}
\end{equation}
In order to get a stable trajectory it must be sufficiently
small. Now, a magnification of $\Delta \tau$ after some time steps
causes a ``coherent'' jump of the whole system out of a local
minimum. This is exactly the desired behavior of the ``Simulated
Annealing'' method. But in that case the ``decoherent'' variation of
the configuration shifts the system configuration to a completely
unrelated point, permitting almost no effort at all in minimizing
$S$. Thus, the system starts again from a worse configuration. It
converges extremely slowly for a considerable number of degrees of
freedom. However, the coherent annealing of our Pseudo-Langevin approach allows
to ``jump'' over local hills without completely loosing properties of
the configuration. It  permits a larger time step to overcome
small bumps and troughs, see Fig.~\ref{fig:bump}.

This method is quite efficient and sometimes called ``gradient
descent'' method. An interesting variant is also used in
\cite{Moore261,MooreServant} which directly deal with the saddle point
configuration.

The limit of our method is reached with derivatives $\frac{\de
S[q]}{\de q_i(n)}$ in (\ref{PL}) for the $q_i$ varying on strongly
differing scales. Then the time step can not be computed appropriately
for all $q_i$ simultaneously and the system runs over a minimum at one
$q_i$ while another $q_j$ has not changed considerably at all. Such a
behavior appears for large $h^3$-bumps in the effective
potential. This may already be the case in the NMSSM for some parameter
combinations. But a combination of all presented methods can also
treat more complicated cases we cannot discuss here.

What can be done to avoid some of the problems and to rate the quality of
a minimum found? First it is important to have an ansatz as close as
possible to the desired solution to avoid reaching an unwanted local
minimum after the time consuming second minimization step. One can
also increase the number of intermediate steps in the procedure to
improve the precision at the cost of computing time.  Additionally we
give an independent check to rate the results.
\paragraph{Rating of solutions}
As long as we have energy conservation, one can check the quality of
the results easily. The relation
\begin{eqnarray}
& &V[q_i]/T[q_i]=1\\
\text{or} & & V[q_i]-T[q_i]=0\quad\text{for all $x$ (or $\tau$)},
\end{eqnarray}
where $T[q_i]$ is the kinetic part of the action, must be fulfilled
and one can check the deviation. With our Pseudo-Langevin approach the
best solutions reach $V[q]/T[q]-1$ with a precision of order
$10^{-3}$.  However, such a high precision is not necessary, even for
sensible quantities like $\Delta\beta$.  One can construct further
identities that have to be fulfilled like an extended virial theorem
as in \cite{Kusenko}.

There is also a possibility for cases without energy conservation. Let
us first discuss the problem of shooting along a given path in the
Higgs field space because this method also gives a quality check for
non energy conserving cases.  Imagine that we have chosen a path for
the bubble wall solution in field space. This does not mean that we
have solved the equations of motion which would yield the $x$
dependence of the fields.  But for a given path we can use the
over\-shoot\-ing-under\-shoot\-ing procedure to find the
$x$-dependence.  The naive method to take the potential along the path
as an effectively 1-dimensional potential gives wrong results even for
cases with energy conservation.  It can lead to a wall thickness which
differs from the real result by a factor of two or more. The reason
for this is that the differential equations only contain partial
derivatives.  In the naive approach the effective derivative (along
the path) is a combinations of the original (partial) derivatives. But
it can be extended to construct a quality check of solutions by
looking at each field space direction separately.

One necessary condition for reliable results is the shooting condition
for the bounce solution: The ``marble'' rolling along an arbitrary
path must come to rest at the zero minimum (symmetric
phase). The field configuration ${\bar q}$ which defines the path
coordinates is a combination of the scalar fields
$(h_1,h_2,\ldots)$. A necessary condition for the solution configuration is
\begin{equation}
\int_{x_{start}}^{x_{end}} dx \frac{\de V[{\bar q}]}{\de h_k}=0,\label{shoot1}
\end{equation}
which has to be fulfilled separately for all field directions
$k$. Each configuration $h_k(x)$ can be projected to the chosen path
$\bar{q}$ leading to configurations $\bar{q}_k(x),\ k=1\ldots N$
for $N$ fields. Vice versa, assuming that $\bar{q}(x)$ is the exact
solution to the (critical) bubble wall equations, all $\bar{q}_k(x)$
are then identical by construction.  We can use this to construct a
general check. The configuration $\bar{q}(x)$, obtained by shooting
along a chosen path, may be projected to all components $h_k(x)$ in
the field space which automatically fulfill (\ref{shoot1})
independently.  The inverse functional $x[h_k]$ can subsequently be
used to construct a motion configuration $\bar{q}$ along the path for
every $k$: $\bar{q}_k(x[h_k])$. For the solution this leads to $N$
identical configurations $\bar{q}_k(x[h_k])$. But every pseudo
solution different from the exact solution configuration leads to
differences in these reconstructed pseudo-solutions which rates the
quality of the pseudo solution.

Concluding, the path has to be chosen that
\begin{equation}
\int_{x_{start}}^{x_{end}} dx \frac{\de V}{\de h_1}\ldots
=\int_{x_{start}}^{x_{end}} dx\frac{\de V}{\de h_i}=\ldots=0\quad\forall h_i
\end{equation}
(shooting condition) {\em and simultaneously} the (inverted) motion
along the path must be the same for every field $h_i$. This even works
for critical bubbles where we have no energy conservation and the
energy check does not work.

The methods presented here can be used to investigate the dynamics of
bubble wall expansion and baryogenesis in more detail. In
Section~\ref{sec:CPgeneral} we extend them to investigate CP violation
for many parameter combinations. Our considerations and calculations
were also essential to develop better perception of which
approximations might be useful and appropriate to simplify
calculations as done in Section~\ref{sec:BAU}. More applications can
be found in refs.~\cite{HuJoSchmiLai,JSwall,HS,HS00}
%
%
%
%
%
\section{CP violation}
\label{sec:CPgeneral}
To calculate the actual baryon asymmetry CP violation as one of the
Zakharov conditions becomes important. In
\cite{HuetNelson,non-eq,CJK,ASO,risa,seco00} it was found that a
sufficient baryon asymmetry requires CP violating phases as large as
${\cal O}(10^{-1})$. This might lead to contradictions with
experimental constraints~\cite{edmphan} on the electric dipole moment
of the neutron (EDM). It is still under discussion whether this
contradiction might be resolved by special cancellations.  Large phases
are also allowed if the 1st and 2nd generation squarks are heavy~\cite{KO}.
Regarding our results in the final Section~\ref{sec:final} this
becomes important again.  For a more detailed discussion of this point
see refs.~\cite{WagnerHiggse} and \cite{edm}.

It is known, that in principle there may be spontaneous CP violation
in the MSSM at zero temperature realized by a CP violating phase
between the two Higgs doublets. While spontaneous CP violation at
$T=0$ is forbidden for the experimental allowed
parameters~\cite{MaPo}, it was suggested that it might be realized at
finite temperature quite naturally~\cite{ComelliPietroni,QuirosCP} or
especially only in the bubble wall between the symmetric and the broken
phase~\cite{cpr,fkot,CKVischer}.

An interesting scenario is therefore a temperature induced CP
violation in the bubble wall during the first order electroweak phase 
transition which is not restricted by {\em any} experimental
bounds. Hence it may be maximal and might sufficiently
support the generation of the baryon asymmetry of the
Universe\cite{JPT1,mstv}. It will be referred to as {\em transitional CP
violation} in contrast to CP violation in the broken minimum. In the
literature~\cite{ComelliPietroni,QuirosCP,cpr,fkot,CKVischer,Mikko3dCP}
the conditions for transitional CP violation were widely discussed. In
Section~\ref{sec:numana} we will revisit this question by following
the approach of ref.~\cite{HuJoSchmiLai} in more detail. Since this
approach is different to previous ones we compare the methods.

Explicit CP violation is introduced by complex mass parameters in the
effective potential. Independently from strong experimental bounds it
might be interesting to study its effects on bubble walls and whether
it suffices to contribute considerably to the baryon asymmetry. In
Sec.~\ref{ssec:explicitM} we will analyze in detail the evolution
of explicit CP violating phases in the bubble wall.

\subsection[Transitional CP Violation in the MSSM]{Transitional CP Violation  
in the MSSM: Revisited}
\label{sec:numana}
In this section we elaborate on the systematic search
of~\cite{HuJoSchmiLai} for a parameter window for transitional CP
violation in the MSSM. First we consider the case without explicit CP
violating phases and check whether a CP conserving solution
($\theta=0,\pi$) is a local minimum of the action.  Instability
in the $\theta$-direction requires
\begin{equation}
m_3^2(h_1,h_2):=\left.\frac{1}{|h_1h_2|}\frac{\partial^2V_T(h_1,h_2,\theta)}{\partial\theta^2}\right|_{\theta=0}<0.
\label{instcond}
\end{equation}
$m_3^2(h_1,h_2)$ is a measure for the instability at a given point in
the wall. $V_T$ is the finite temperature effective potential
including stops, charginos and neutralinos.  In our convention $h_1$
can have either sign which permits to consider only
$\theta=0$. Condition (\ref{instcond}) agrees with the constraint of
Lee \cite{Lee} on which \cite{MaPo,QuirosCP,cpr,fkot} are based on. But
Eq.~(\ref{instcond}) is true more generally along the whole bubble wall.

We are interested in regions where (\ref{instcond}) is fulfilled.  In
our search we first neglect the strong experimental bounds on the
Higgs mass since it might still be interesting to find a shape of such a
transitional CP violation.  The region of generically allowed
parameters is an 8-dimensional ``hypercube'' stretched by the tree
level parameters $m_A$ and $\tan\!\beta$, the stop mass parameters
$m_Q$, $m_U$, trilinear couplings $A_t $, and $\mu$ as well as the
gaugino mass parameters $M_1$ and $M_2$ in the neutralino and chargino
mass eigenvalues.  For each chosen parameter set we must determine the
critical bubble wall profile $h_1(z)$, $h_2(z)$ including the
$\theta(z)$ field and scan the wall for points which fulfill the
instability condition (\ref{instcond}).

The methods for solving the (critical) bubble equations with more than
one scalar field including $\theta$ were discussed in
Section~\ref{sec:howto} and in \cite{John3}. However, all these
methods are very time consuming and need sophisticated algorithms.
Therefore we proceed in two steps. First we will restrict the possible
parameter space in order to obtain a more promising parameter set
permitting transitional CP violation. Hence we will investigate the
averaged dependence on the parameters. Second, with this preparation
we will search for a definite parameter set permitting the desired CP
violation in the bubble wall.

A first and useful simplification is to refrain from determining the
wall profile very accurately for each parameter set.  Indeed, in the
MSSM the solution to the equations of motion is quite close to a kink
ansatz $h(z)\sim 1/2(1+\tanh(z/L))$, and therefore we do not need the
full solution to the equations of motion. The solution can be
approximated by the ridge of the negative potential
$-V_{eff}$. Actually, it is even sufficient to approximate the
solution in field space $h_{1,2}$ by the straight line between the
minima.  This approach is correct as long as there is only small CP
violation which does not have a too strong back-reaction to the wall
shape as could happen for large phases (see ref.~\cite{John3}). If
we found a candidate for transitional CP violation with a large phase,
we would have to solve the whole set of equations. But for the moment
we are only searching for promising parameter regions.

In order to perform a complete search for a parameter window with
transitional CP violation in the bubble wall it is important to
consider the dependences on the various parameters.  At tree level we
find
\begin{equation}
m_{12}^2=-\frac{1}{2}m_A^2\sin\!2\beta.
\end{equation}
Hence, the minimum in $\theta$ direction is at $\theta=0$. Thus, in
order to obtain spontaneous CP violation the 1-loop contributions have
to overcome the tree level term. Obviously large $\tan\!\beta$ and
small $m_A$ are preferred. The dependence on other parameters is not
quite obvious. Potentially it is possible that $m_3^2$ is negative for
very small (negative) $m_U^2 << (2\pi T)^2 << m_Q^2$
\cite{HuJoSchmiLai,Mikko3dCP}.

\begin{figure}[h]
\vspace*{1cm}
\hspace*{.0cm}
\begin{picture}(0,200)
\put(20,0){\epsfysize=8cm\epsffile{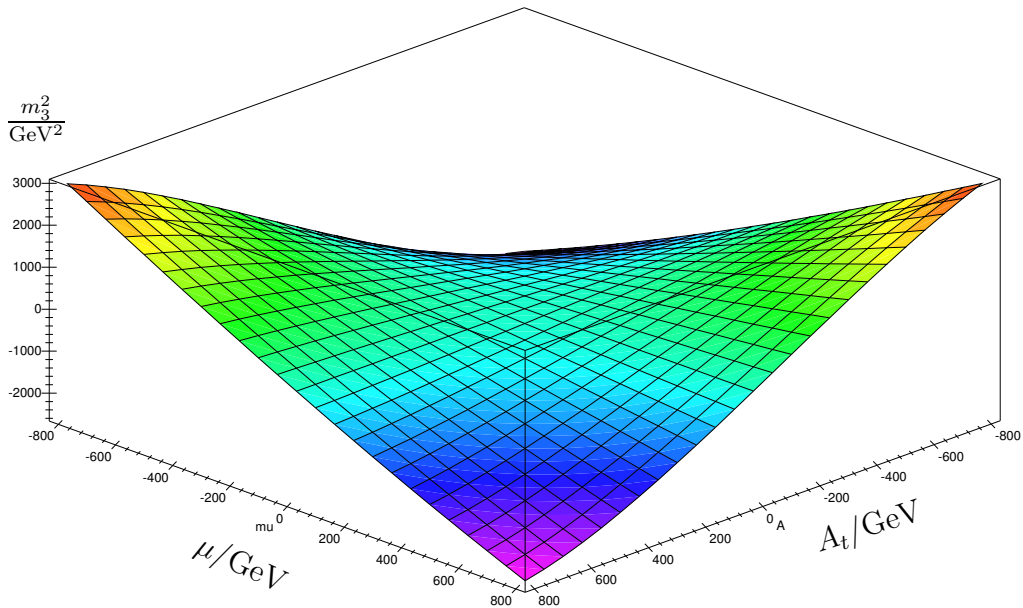}}
\put(20,180){$\frac{m_3^2}{\text{GeV}^2}$}
\put(90,20){\rotatebox{335}{$\mu$/GeV}}
\put(325,20){\rotatebox{18}{$A_t$/GeV}}
\end{picture}
\vspace*{.5cm}
\caption{Dominant 1-loop effect: $m_3$ in dependence of  $A_t$ and
$\mu$ including stops and charginos. The tree level contribution causes a
shift in the mean $m_3^2$. Large values with equal signs are preferred to give  
a negative
contribution to $m_3^2$.}
\label{fig:mapleatmu}
\end{figure}

In an analytical approach where we plot $m_3^2$ versus various
parameters we always fix the remaining parameters. We find that it
dominantly depends on $m_A$ and $\tan\!\beta$.  All other parameters are
effectively 1-loop corrections and give a correspondingly smaller
effect. The strongest 1-loop dependence is given by $A_t$ and $\mu$
(see Fig.~\ref{fig:mapleatmu}).  We also find here that large values
with opposite signs give the smallest $m_3^2$. Later we will see that
the latter effect appears only to be correct in these preliminary
analytical investigations.

An interesting result we obtain by this method is that
$m_3^2(h_1,h_2)$ tends to become larger while approaching the broken
phase within the wall. The opposite behavior was noticed in
\cite{fkot}. An increasing $m_3^2$ is undesired since large phases in
the broken minimum are restricted by the experiments. Also small
phases are excluded by experimental bounds on the Higgs masses which
are connected with small CP violation \cite{MaPo}. So even if a
transitional CP phase occurs analytically it might be excluded by
experiments.

The other parameters give minor effects which nevertheless can
contribute.  Large $M_2$ and $\mu$ are favored in order to get small
$m_3^2$.  As already mentioned, in this kind of analysis we always fix 
most of the parameter and vary only some of them. But we are
interested in the whole parameter space. We will see that the shown
dependences must be treated with caution since they may change using
more general investigation methods.

In the next stage we want to investigate the complete parameter space
in a wide range. We therefore choose a different strategy. We will
analyses correlations in average.
At the first stage, we do not solve for $h_1,h_2,T_c$, but rather
take them as free parameters in the ranges
$h_1/T=-2...2$ and $h_2/T=0...2$, $T=80...120$ GeV.
The zero temperature parameters are varied in the wide ranges
\begin{eqnarray}
& & \tan\!\beta  = 2...20, \qquad m_A = 0...400 \mbox{ GeV}, \nonumber\\
& & m_U =  -50...800 \mbox{ GeV}, \qquad m_Q = 50...800 \mbox{ GeV}, \label{paspa}\\
& & \mu, A_t, M_1, M_2 = -800...800  \mbox{ GeV}. \nonumber
\end{eqnarray}
Here a negative $m_U$ means in fact a negative
right-handed stop mass parameter, $-|m_U^2|$. We have also
studied separately the (dangerous~\cite{MooreServant}) region where
the transition is very strong~\cite{2loop,litestop}
corresponding to $m_U \sim -70...-50 \mbox{ GeV}$. 

Note that since we do not solve for the equations of motion at this
stage but allow for $h_1=\pm |h_1|$, we have to divide in
(\ref{instcond}) by $h_1h_2$ instead of $|h_1h_2|$: this leads in
general to positive values due to the tree-level form of the
potential. A signal of a potentially promising region is then a small
absolute value of the result, since this means that we are close to a
point where $\partial_\theta^2 V_T(h_1,h_2,\theta)$ crosses zero.

\begin{figure}[h]
\vspace*{1cm}
\hspace*{0.3cm}
\centerline{\epsfysize=9.6cm\rotate[r]{\epsffile{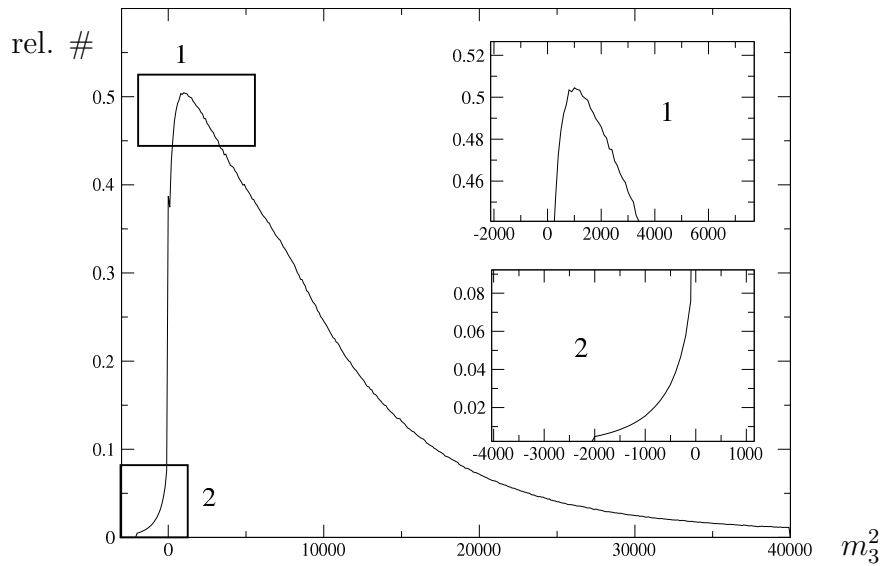}}}
\caption{Relative number versus value of $m_3^2$. Small positive values are  
preferred.}
\label{fig:nom3}
\begin{picture}(0,0)
\put(360,40){ $m_3^2$}
\put(50,230){rel. \#}
\end{picture}
\end{figure}

In Fig.~\ref{fig:nom3} the relative number of $m_3^2$ versus its
occurring values demonstrates which values are the most likely ones.
There is a strong peak at a certain positive value and the probability
decreases rapidly for zero or negative values of $m_3^2$. But
nevertheless there are some negative $m_3^2$. It has to be checked,
whether they are relevant for the bubble wall.

We found that spontaneous CP violation is statistically independent of
the actual critical temperature. For checking the individual cases it
may nevertheless play a role \cite{Mikko3dCP,HuJoSchmiLai}.

\begin{figure}[h]
\centerline{\epsfysize=10cm\epsffile{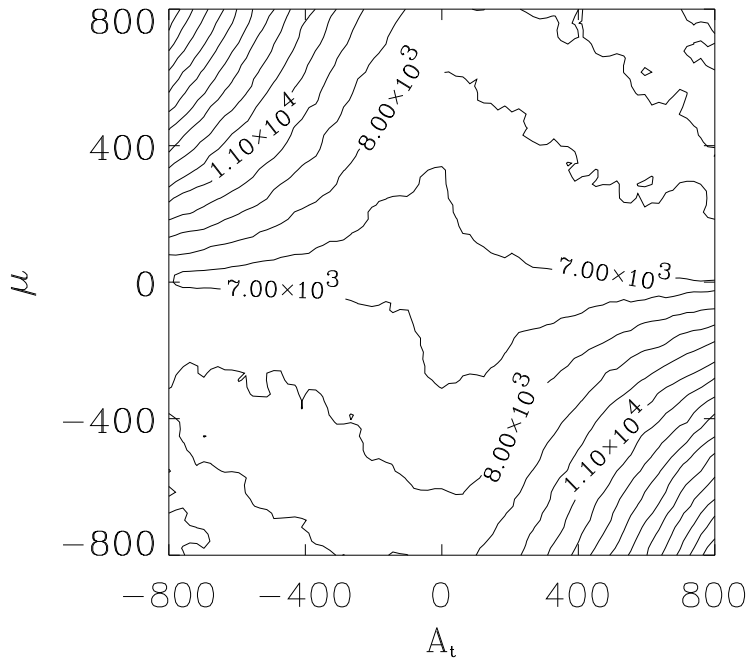}}
\vspace*{-0.5cm}
\caption{A totality of $1.03\cdot 10^9$ averaged $m_3^2$ versus $A_t $ and  
$\mu$. The favorite region is
the cross like region around vanishing $A_t $ and $\mu$.}
\label{fig:3dAtmu}
\end{figure}

In the analytical result we observed that large $\mu$ and $A_t $ with
equal signs are preferred. This can be checked in a two dimensional
correlation plot in Fig.~\ref{fig:3dAtmu}. Both combinations are
correlated with large positive $m_3^2$, hence with in average more
stable regions. The best region is a cross-like region where either
$A_t $ or $\mu$ are small.  This could not be expected from the
analytical approach which showed a preference of large values of
$A_t$ and $\mu$.
\begin{figure}[h]
\vspace*{1cm}
\centerline{\epsfysize=10cm\rotate[r]{\epsffile{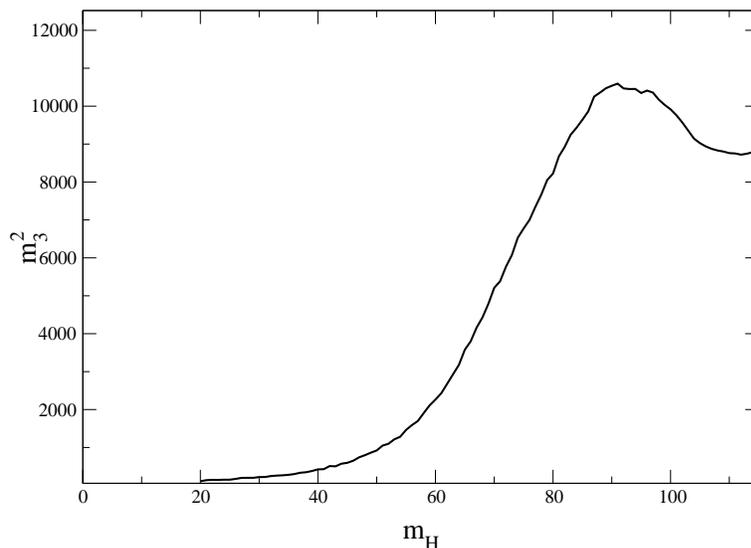}}}
\vspace*{-.5cm}
\caption{Correlation of $m_3^2$ and the Higgs mass $m_H$. Most favorable
values are correlated with unphysical small Higgs values.}
\label{fig:m3_mH}
\end{figure}

The strongest correlations come from $A_t$ and $\mu$. The scale of
their variation is larger by a factor of about 3 than the scales of
the 1-loop mass parameter dependences with averaged 1-loop parameters.

Unfortunately small $m_A$ and large $\tan\!\beta$ are strongly preferred
to obtain an unstable CP direction. This is in contradiction to the
requirements of a strong first order phase transition
\cite{CarlosEspinosa}. Moreover CP odd Higgs masses up to $m_A\sim
100$~GeV are experimentally excluded.

In \cite{MaPo} it was argued that spontaneous CP violation at $T=0$ is
excluded because of the experimental exclusion of the corresponding
small Higgs masses. Accordingly, for transitional CP violation
physical Higgs masses lower than around $40$GeV are fairly probable,
see Fig.~\ref{fig:m3_mH}. Here the average of $m_3^2$ decreases
rapidly.

\begin{figure}[h]
\epsfysize=6cm{\epsffile{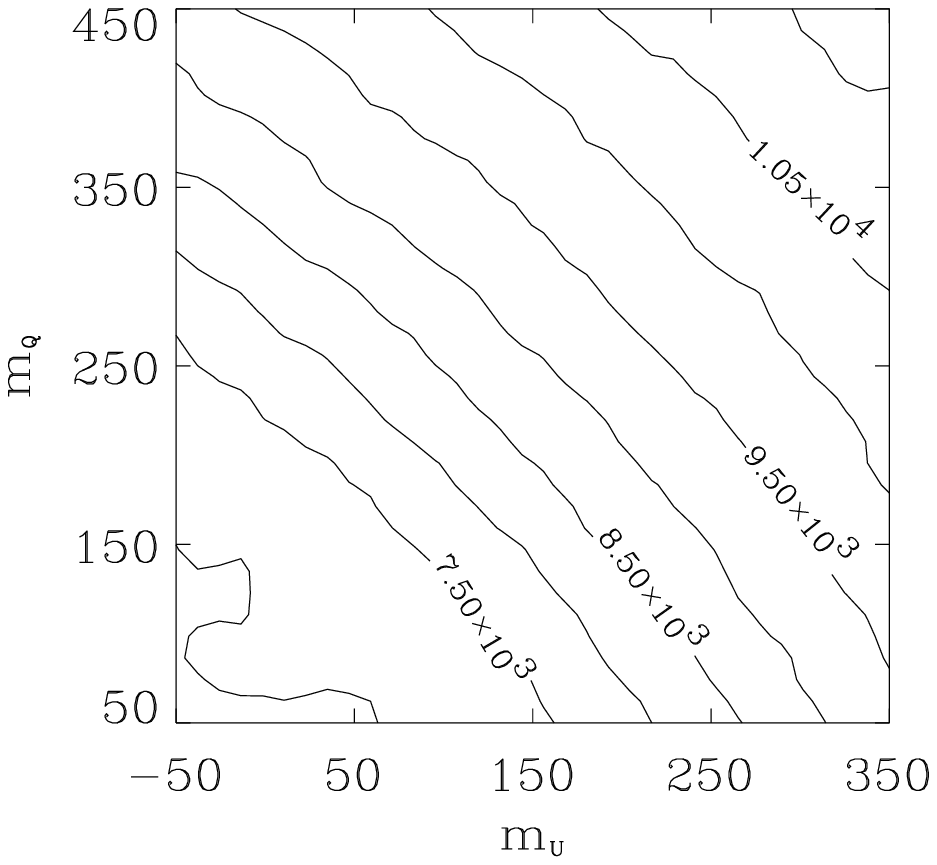}}
\epsfysize=6cm{\epsffile{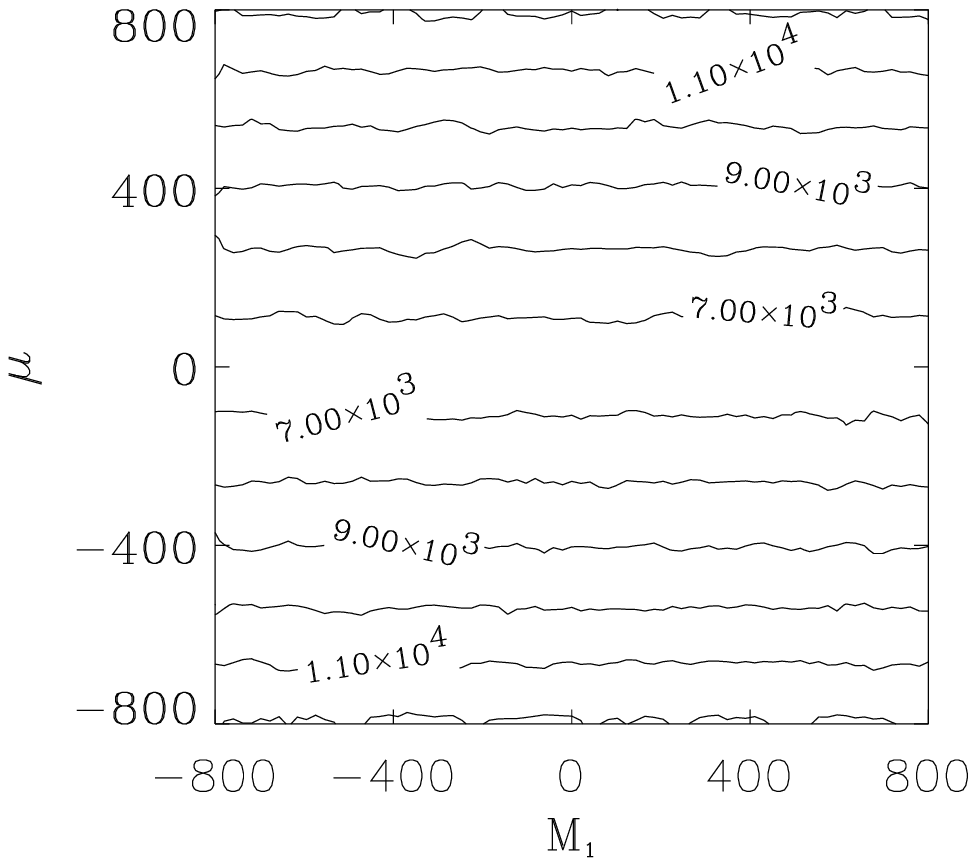}}\\
\epsfysize=6cm{\epsffile{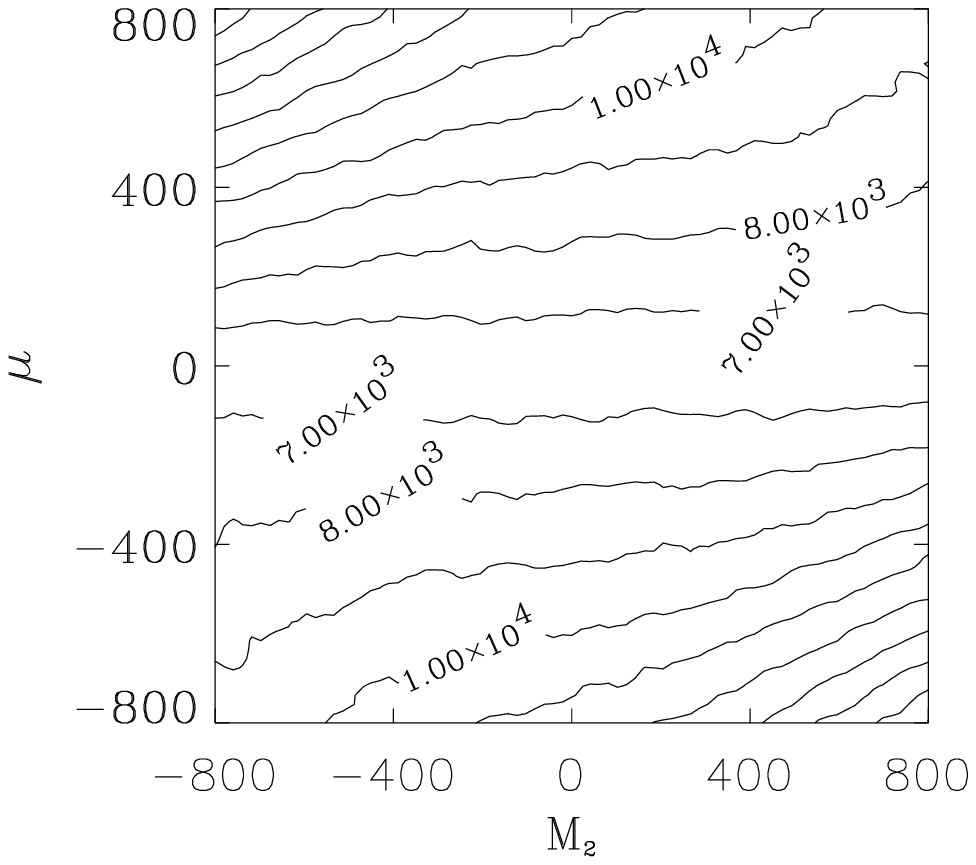}}
\epsfysize=6cm{\epsffile{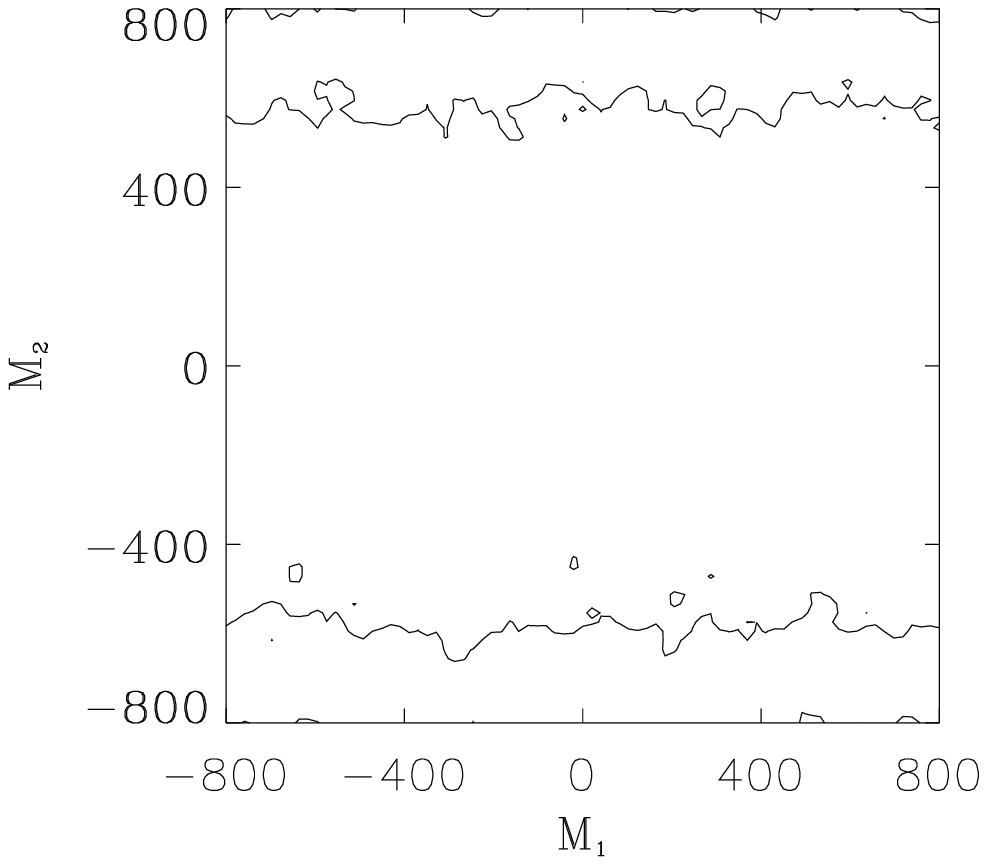}}
\caption{Correlations of the 1-loop parameters in average. The measure $m_3^2$ for 
instability versus the average of the other parameters is shown. The
strongest correlation appears for the mass parameter $\mu$. $M_1$ and
$M_2$ show less average dependences. Small $m_U$ are preferred,
which coincides with the requirements of a strong phase transition
whereas also small $m_Q$ are preferred, which points into the wrong
direction.}
\label{fig:rest}
\end{figure}

There are also minor effects from the remaining parameters. The stop
parameters $m_U$ and $m_Q$ are preferred to be small
(Fig.~\ref{fig:rest}a). The effect is smaller than the main 1-loop
contribution from $A_t $ and $\mu$. $M_1$ and $M_2$ are almost
completely uncorrelated. There are smaller correlations with $A_t $ or
$\mu$ (Figs.~\ref{fig:rest}b,c) which are folded into a vanishing
correlation of $m_3^2$ with $M_1$, $M_2$ (Fig.~\ref{fig:rest}d).  The
effects on $m_U$, $m_Q$, $M_1$, and $M_2$ do not strongly promote a
small or negative $m_3^2$, but nevertheless they can be taken as  hints.

In a second stage, we study the relevant parameter region in more
detail. We start from the full parameter range (\ref{paspa}) and
restrict the search to the most favorable region. First, we determine
the critical temperature. Then the equations of motion are solved for
$(h_1,h_2)$. By comparing with the numerical exact results in several
cases we find that a sufficient accuracy can be obtained by searching
the ``ridge'' as an approximation, see Section~\ref{sec:howto}. Finally,
we look for the minimum in $\theta$ at fixed $(h_1,h_2)$ which is a
fast and reliable approximation to the real solution as long as
$\theta$ is small. We make some further restrictions: we exclude
unphysical negative mass parameters as $m^2_{\tilde t}<0$ and
discarded extremely weak phase transitions $v/T\ll1$. We also excluded
cases leading to $T=0$ spontaneous CP violation in the broken phase
since this requires very small values of $m_A$\cite{MaPo}.  

While in the statistical analysis we included $10^9$ parameter sets,
we investigated $\sim 2\times 10^6$ individual configurations in more detail.
Nevertheless, we could not find any case with the desired property of
temperature induced transitional CP violation within the corresponding
bubble wall in the MSSM. Furthermore, in \cite{fkot} a special point
around $m_U^2\approx0$ was considered. Since thermal mass were
neglected, this corresponds to the point were $m_U^2+\#T^2\approx
0$. This point is in the vicinity of charge and color breaking
minima. Without expanding the 1-loop contribution in $v_1/v_2$, we
cannot reproduce the instable behavior proposed in \cite{fkot}.

Moreover we found that in the MSSM the dominating effects which
support small $m_3^2$ are in contradiction to strong phase transitions
and baryogenesis requirements.
\subsection{Explicit CP Violation}
\label{ssec:explicitM}
We investigated also explicit CP violation in the wall. Although there
are strong experimental constraints on the magnitude of the
corresponding phases it might be interesting to see how strong the
variation of $\theta$ along the phase boundary can be, and if it
contributes considerably to the baryon asymmetry.

There are several possibilities to include explicit CP violation in
the effective potential. In principle, several mass parameters might be
complex.  At tree level this is $m^2_{12}$, and at 1-loop level we
have more possibilities through the masses of stop, chargino, and
neutralino.  In the stop, chargino, and neutralino sector $\mu$, $A_t$
and the soft breaking mass terms can carry phases:
\begin{eqnarray}
\theta_{m_{12}} \leftrightarrow m_{12}^2=\mu B,& \theta_{A_t} \leftrightarrow  
A_t,& \theta_{\mu} \leftrightarrow \mu, \nonumber\\
\theta_{M_1} \leftrightarrow M_1, & \theta_{M_2} \leftrightarrow M_2&.
\end{eqnarray}
Altogether we have five phases. $\theta_{m_{12}}$ can be absorbed in
the Higgs fields and then arises in the mass matrices. The phases in
the stop off diagonal entries can be absorbed into a complex ${\tilde
A}_t=A_t -\mu\cot\beta$.  We define the remaining three phases as
$\theta_A=\theta_{m_{12}}+\theta_{A_t}+\theta_\mu$,
$\theta_C=\theta_{m_{12}}+\theta_\mu+\theta_{M_2}$, and
$\theta_M=\theta_{m_{12}}+\theta_\mu+\theta_{M_1}$ where
$\theta_A$ appears in the stop matrix, $\theta_C$ in the chargino
matrix, and together with $\theta_{M_1}$ in the neutralino matrix. In
the chargino and stop eigenvalues we then have to replace $\cos\theta
\to \cos(\theta+\theta_A)$ and $\cos\theta \to \cos(\theta+\theta_C)$.
The replacement in the neutralino eigenvalues is more complicated and
will not be given here.

\begin{figure}[h]
\vspace*{.5cm}
\centerline{\epsfysize=10cm\rotate[r]{\epsffile{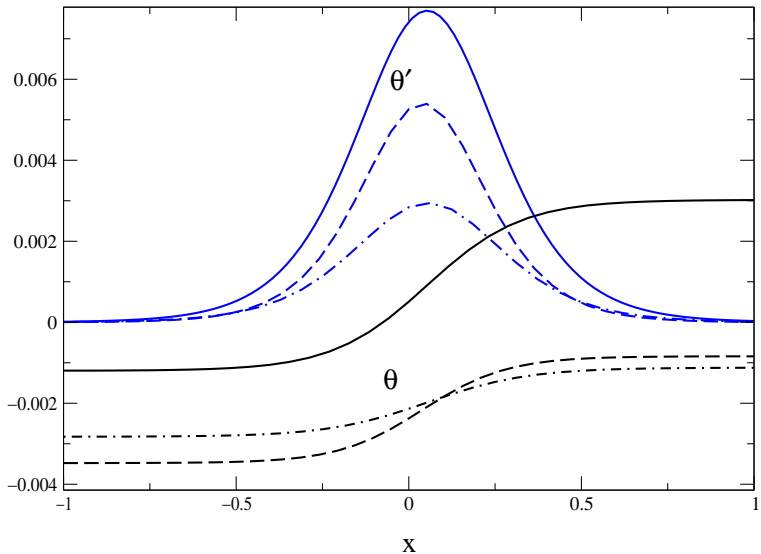}}}
\caption{Dependence of $\theta$ and $\theta'$ on $m_A$ near
physically realistic cases: The derivative strongly depends on $m_A$
and $\tan\!\beta$. Solid: $m_A=80GeV$, $\tan\!\beta=2.0$, dashed:
$m_A=120GeV$, $\tan\!\beta=2.0$, dot-dashed: $m_A=120GeV$,
$\tan\!\beta=3.0$. ($x$ and $\theta'$ in units of $GeV^{-1}$.)}
\label{explicit2}
\end{figure}

Again we investigate the variation of $\theta$ along the wall.  As in
the spontaneous case, we search for minima along the $\theta$
direction. For baryogenesis, we need a strong variation of the phase
$\de_x\theta(x)$ along the wall. But even for maximal angles $\theta_A=\pi/2$,
$\theta_C=\pi/2$, and $\theta_{M_1}=0$ we find a strongly suppressed
CP phase in the broken Higgs phase.  For $m_A \lsim 90GeV$ $\theta(x)$
is of order $10^{-3}\ldots 10^{-2}$ and varies only moderately within
the wall as shown in Fig.~\ref{explicit2}.  Especially there is no
kind of peak as proposed earlier \cite{fkot}. Only for experimentally
excluded values of $m_A\lsim$~10GeV we do obtain phases up to order
unity. For explicit phases of order $10^{-1}$ the dynamical phase is
typically of order $10^{-4}$. We again realize that the size of the variation 
is dominated by the tree level parameters $m_A$ and $\tan\!\beta$.

From a technical point of view the evolution of the phase $\theta(x)$
into the symmetric phase is of interest. It is more difficult to
determine (and less meaningful). The solutions of Fig.~\ref{explicit2} have
been obtained by using a $tanh$-ansatz also for $\theta(x)$, which
turned out to compare very well with the numerical precision solution in
the middle of the wall, where the full equations of motion can be
solved more easily.
\subsection{CP Violation in the NMSSM}
\label{sec:NMSSMCP}
In the NMSSM we have many possibilities and a much larger parameter
space \cite{HS}. Besides the Higgs fields $h_1$
and $h_2$ we have a singlet field $S$. Permitting complex fields, we
can write the singlet field $S=n+ic$ implicitly introducing a phase
$\theta_s=\mathop{\arctan}(c/n)$.  Indeed, it turns out that
transitional CP violation can occur in the NMSSM quite naturally.
In \cite{HuJoSchmiLai} an example is shown where in the $SU(2)\times
U(1)$-symmetric phase the singlet carries a vev in both components:
$(n,c)_{\text{symm.}}=(50.4GeV,99GeV)$ and
$(n,c)_{\text{broken}}=(-122GeV,0GeV)$. Moreover the standard Higgs
phase varies between zero and a value of order $\theta_{sym}\approx
1/20$ (in this special example) in the symmetric phase and zero in the
broken phase. Both phases, $\theta$ and $\theta_s$ start with
non-vanishing values in the symmetric phase and vanish in the broken
phase.

We used the values $k=0.4$, $\lambda=0.05$, $\mu=212.6$GeV,
$m_Q=278.1$GeV, $m_U=209.0$GeV, $B=83.9$, $r=533.2$, $\tan\!\beta= -5.0$,
$A_t=-219.8$GeV, $A_k=50$GeV, $A_\lambda=-150$GeV, $M_1=41.3$GeV,
 $M_2=82.2$GeV \cite{HS00,HuJoSchmiLai}. The critical temperature is
$T_c=102.2$GeV.

This demonstrates the general possibility of transitional CP violation
in bubble walls of the NMSSM phase transition. We have a strong phase
transition in the example case. This shows the principle viability of
baryogenesis in the NMSSM. More detailed calculations can be found in
\cite{HS00}.
%
%
%
%
%
%
\section{The Baryon asymmetry in the MSSM in the semi-classical limit}
\label{sec:BAU}
\subsection{Generalities}
As just discussed there can be explicit CP violation in the MSSM and
also transitional CP violation in the NMSSM. The expanding bubble walls
of a first order phase transition sweeping through the hot plasma
generate temporary non-equilibrium. In the hot plasma we have sizeable
baryon number violation by the weak sphaleron which rapidly stops in
the equilibrium Higgs phase. Thus baryogenesis in principle is
possible.  However, we have to describe a concrete procedure since
these are only necessary conditions. Technically one has to study
transport equations in the plasma for particles/fields with varying
mass due to the varying Higgs fields in the bubble wall. For thermal
momenta $p\sim T$ and a ``thick'' wall ($L_w\gg 1/T$) the
semi-classical approximation should be appropriate and we should end
up with Boltzmann transport equations. In principle one has to start
with the Baym-Kadanoff equations of the real time non-equilibrium
formalism and end up with classical particle equations (plus some
corrections as memory effects). In a different corner for soft bosonic
fields a classical field description is expected.  Here we only will
deal with the particle limit and discuss the WKB approximation in
accordance with refs.~\cite{CJK,CJKneu}. Since we are mainly
interested in the baryon asymmetry we will first calculate the transport
for an asymmetry between left-handed particles and their CP conjugates
and then consider baryogenesis via (out of equilibrium) sphaleron
transitions in the symmetric phase. The transport equations for
particles alone are interesting if the friction of the bubble wall and
the resulting stationary wall velocity is calculated. Here CP
violation can be neglected.  As it will turn out that the wall
velocity strongly influences baryogenesis this is an important point,
not being dealt with here, but we refer to
refs.~\cite{MP12,JSwall,GMoo,ariel}.
\subsection{WKB approximation and dispersion relations}
In the following we summarize the derivation of the chargino
dispersion relations in the presence of a slowly varying Higgs field
background.  For the details of the calculation we refer the reader to
ref.~\cite{HS00} where also the application to squarks was worked out. Indeed
the contribution of squarks turned out to be sub-leading.

To obtain the dispersion relation we solve the Dirac equation for the
charginos in the WKB approximation
\begin{equation}\label{fwkb1}
 \left(\begin{array}{cc}-M &
                      i\sigma^{\mu} \partial_{\mu} \\  i\bar\sigma^{\mu}
                      \partial_{\mu}  & -M^{\dagger}\end{array}\right)
                {\Psi_L \choose \Psi_R}=0,
\end{equation}
where
\begin{equation} \label{char_mass}
M=\left(\begin{array}{cc}
M_2&g_2(H_2^0)^* \\ g_2(H_1^0)^* &\mu
\end{array}\right)
\end{equation}
and $\Psi_L=(\tilde W^+,\tilde h_2^+)^T$, $\Psi_R=(\overline{\tilde
W^-},\overline{\tilde h_1^-})^T$.  Since spontaneous CP violation does
not occur in the MSSM, we have to rely on explicit CP violation by the
$M_2$ and $\mu$ parameters.  At this level all interactions between
the particle and the plasma are neglected. The scattering effects will
be accounted for in Section 5.3 when Boltzmann equations are written down
to describe the local phase space distributions.

During the passage of the bubble wall the chargino mass matrix becomes
space-time dependent.  We neglect the curvature of the wall and boost
to the rest frame of the bubble wall. Then the mass matrix only
depends on the position relative to the wall, i.e.~$M=M(z)$.

Exploiting conservation of energy and boosting to
the Lorentz frame where the momentum parallel to the wall
vanishes, we take the ansatz
$\Psi=e^{-iEt}\xi(z)$.
Eq.~(\ref{fwkb1}) separates into two equations \cite{JPT1}
\begin{equation} \label{fwkb2}
i\partial_z\xi_{\pm}=\pm Q(z)\xi_{\pm}, \qquad Q(z)=\left(\begin{array}{cc} E&
-M^{\dagger}(z) \\ M(z) &-E
\end{array}\right)
\end{equation}
where $\xi_{\pm}$ are the $S_z=\pm\frac{1}{2}$ components of $\xi$.
To solve Eq.~(\ref{fwkb2}) one brings the $z$-dependent matrix
$Q(z)=D(z)Q_D(z)D(z)^{-1}$ into a diagonal form, where \cite{HS00}
\begin{eqnarray} \label{fwkb2a}
Q_D&=&\left(\begin{array}{cc} \sqrt{E^2-m^2}&0\\ 0 & -\sqrt{E^2-m^2}
\end{array}\right),
\\[.2cm] \nonumber
D&=&\left(\begin{array}{cc} U & 0 \\ 0 & V \end{array}\right)
\left(\begin{array}{cc} \cosh X&e^{-i\theta}\sinh X\\
e^{i\theta}\sinh X &\cosh X \end{array}\right).
\end{eqnarray}
The SU(2) matrices
\begin{equation}  \label{fwkb11}
{U}=
\left(\begin{array}{cc} \cos a & e^{-i\gamma}\sin a \\
  -e^{i\gamma}\sin a & \cos a
\end{array}\right), \qquad
{V}=
\left(\begin{array}{cc} \cos b & e^{-i\delta}\sin b \\
  -e^{i\delta}\sin b & \cos b   \end{array}\right)
\end{equation}
are determined by $M=VM_{D}U^{\dagger}$, where $M_D={\rm
diag}(m_1e^{i\theta_1},m_2e^{i\theta_2})$.  The entries of $X$ which
is a diagonal $2\times2$ matrix in flavor space, obey $\tanh
2X_{1,2}=m_{1,2}/E$.  In the local WKB basis
$\tilde\xi_{\pm}=D^{-1}\xi_{\pm}$ the Dirac equation (\ref{fwkb2})
takes the form
\begin{equation} \label{fwkb3}
i\hbar\partial_z\tilde\xi_{\pm}=(\pm Q_D-D^{-1}i\hbar\partial_zD)
\tilde\xi_{\pm}
\end{equation}
which still is an exact equation. For mixing Dirac fermions like the
charginos the transformation to the WKB basis consists of a rotation
in flavor space (first factor in $D$) and a rotation to the helicity
basis (second factor in $D$). In case of a single Dirac fermion,
e.g.~the top quark, there is only the helicity rotation
\cite{JPT1,HS00}, while for scalar particles, e.g.~the stops, only the
flavor rotation is present \cite{HS00}.

In general, the correction term $D^{-1}i\hbar\partial_zD$ caused by
the position dependent field redefinition is not of diagonal form. The
components of $\tilde\xi_{\pm}$ are still coupled. However, the
off-diagonal part is proportional to $\partial_z D\sim D/L_w$. Typical
momenta of the particles in the plasma are of the order of the
temperature $T$, which is much larger than $1/L_w$ for the bubbles
under consideration. We therefore expand Eq.~(\ref{fwkb3}) in powers
of $\partial_z$ or more precisely in powers of $\hbar$ (WKB
approximation) that we already reintroduced for that reason.

To order $(\hbar)^0$ we can neglect the $D^{-1}\hbar i\partial_z D$
contribution. Thus the components of $\tilde\xi_{\pm}$ decouple in
Eq.~(\ref{fwkb3}).  Inserting the WKB ansatz for the fermion field
\begin{equation} \label{wkb0}
\tilde\xi_{\pm}\sim e^{-\frac{i}{\hbar}\int^zp_z(z')dz'}
\end{equation}
into (\ref{fwkb3}), we obtain the dispersion relations
\begin{equation} \label{fwkb0a}
p_z(E)={\rm sgn}(p_z)\sqrt{E^2-m^2}.
\end{equation}
Obviously, in the classical limit the $\theta$-dependence completely
disappears, demonstrating that CP violation is indeed a
quantum-mechanical phenomenon.

To solve the Dirac equation (\ref{fwkb3}) to order $\hbar$ we have to
take into account the $D^{-1}\hbar i\partial_z D$ term which
reintroduces a coupling between the two components of
$\tilde\xi_{\pm}$. The dispersion relations $p_z(E)$ are obtained from
the eigenvalues of the matrix $\pm Q_D-D^{-1}i\hbar\partial_zD$.
Since to order $\hbar$ the off-diagonal terms do not contribute we are
left with \cite{HS00,CJK}
\begin{eqnarray} \label{fwkb13}
L_I:\mbox{ }
p_z=\mbox{sgn}(p_z)\sqrt{E^2-m^2_I}-(\theta_I'+\delta'\sin^2 b)
\sinh^2 X_I+\gamma'\sin^2 a\cosh^2 X_I, \mbox{ }\nonumber \\
\bar L_I:\mbox{ }
p_z=\mbox{sgn}(p_z)\sqrt{E^2-m^2_I}+(\theta_I'+\delta'\sin^2 b)
\sinh^2 X_I-\gamma'\sin^2 a\cosh^2 X_I, \mbox{ }\nonumber \\
R_I:\mbox{ }
p_z=\mbox{sgn}(p_z)\sqrt{E^2-m^2_I}+(\theta_I'-\gamma'\sin^2 a)
\sinh^2 X_I+\delta'\sin^2 b\cosh^2 X_I, \mbox{ }\nonumber \\
\bar R_I:\mbox{ }
p_z=\mbox{sgn}(p_z)\sqrt{E^2-m^2_I}-(\theta_I'-\gamma'\sin^2 a)
\sinh^2 X_I-\delta'\sin^2 b\cosh^2 X_I, \mbox{ }
\end{eqnarray}
where $I=1,2$ and $\theta_I'=\partial_z \theta_I$, etc.
and
\begin{equation}\label{5.9'}
\sinh^2 X_I=(E-\sqrt{E^2-m^2})/2\sqrt{E^2-m^2}.
\end{equation}
In the symmetric phase $L_2$ and $\bar R_2$ evolve to the left-handed
Higgsinos states $\tilde h_2^+$ and $\tilde h_1^-$, respectively.  The
flavor transformations $U$ and $V$ are related to the parameters of
the chargino mass matrix (\ref{char_mass}):
\begin{eqnarray} \label{fwkb14}
&&\sin^2a=2|A|^2/\Lambda(\Lambda+\Delta)\quad {\rm with}
\nonumber\\
&&A=g_2((M_2H_2^0)^*+\mu H^0_1)\nonumber\\
&&\Delta=|M_2|^2-|\mu|^2+g^2_2(|H^0_1|^2-|H_2^0|^2)\nonumber\\
&&\Lambda=(\Delta^2+4|A|^2)^{1/2}
\end{eqnarray}
and $\gamma=\rm{arg} (A)$. This gives
\[\gamma'\sin^2a=2 Im(A^*A')/\Lambda(\Lambda+\Delta)\]
and there are similar relations for $\sin^2b, \delta$
and $\delta'\sin^2b$ exchanging $a$ and $b$, $H^0_1$ and $H^0_2$,
$\gamma$ and $-\delta$.
The mass eigenvalues (in non-symmetric notation) read
\begin{eqnarray} \label{fwkb15}
[M_D]_{11}&=&M_2\frac{\cos a}{\cos b}-g_2(H_2^0)^*
\frac{\sin a}{\cos b}e^{-i\gamma},
\nonumber\\ { }
[M_D]_{22}&=&\mu\frac{\cos a}{\cos b}+g_2(H_1^0)^*
\frac{\sin a}{\cos b}e^{i\gamma}.
\end{eqnarray}
Notice that the variation of $m_I$, which is encoded in $\partial_z
X_I$, $\partial_z a$ and $\partial_z b$ drops in the dispersion
relations.  The CP violating part of the dispersion relation is
proportional to the {\em derivative} of the phases $\theta_I$,
$\gamma$ and $\delta$.  Thus only varying phases contribute to
CP violation in the semi-classical limit.  Furthermore, CP violation
is proportional to derivatives which guarantees that its effect is
turned off far away from the bubble wall.  Because of the different
dispersion relations, left- and right-handed particles feel a
different (semi-classical) force in their interaction with the
wall. For the anti-particles, $\bar L$ and $\bar R$, the CP violating
part comes with the opposite sign.

As pointed out very recently in ref.~\cite{CJKneu} one should use the
kinetic momentum $p_{\rm kin}=mV_{\rm group}=m\frac{\partial
E}{\partial p}$ instead of the canonical momentum $p$ (we used up to
now) in the classical limit leading to Boltzmann equations, quite in
the spirit of the correspondence principle of basic quantum
mechanics. Calculating the (inverse of the) group velocity from
(\ref{fwkb13}) $E$-independent terms drop out, i.e. $\cosh^2 X_I$ can
be substituted by $\sinh^2 X_I$. The kinetic momenta beyond the zeroth
order (\ref{fwkb0a}) then contain correction terms
$\pm(\theta'_I+\delta'\sin^2 b-\gamma'\sin^2
a){m^2}/({2E(E^2-m^2)^{1/2}})$.  Thus indeed kinetic and canonical
momentum are not equal because of CP violating effects.

The dispersion relations can be inverted. To leading order in the
derivatives the CP violating part of the dispersion relation for the
eigenstate $L_2$ which in the symmetric phase corresponds to $\tilde
h^+_2$ is
\begin{equation}
\Delta E=-{\rm sgn} (p_z)(\theta'_2+\delta'\sin^2 b-\gamma'\sin^2 a)m^2
/2 (p^2_{\rm kin}+m^2).
\end{equation}
(For $p_{\rm kin}\gg m$ this is twice the result one would obtain
with the canonical momentum after substituting $\cosh^2$ by $\sinh^2$ in
(\ref{fwkb13}).) Most importantly $\Delta E$ is now totally symmetric
under the exchange of $H_1$ and $H_2$. This will destroy the most
prominent source term $\sim H_1 H'_2-H'_1H_2$ of older work.

The phases $\gamma$ and $\delta$ only vary due to a change in the
Higgs vev ratio $\tan\!\beta$ or because of transitional CP violation in
the bubble wall.  The first contribution is suppressed, since the
variation of $\beta$ is at most $\sim 10^{-2}$ for realistic Higgs
masses \cite{SecoNum,John3}, while transitional CP violation most
probably does not occur at all in the MSSM as discussed in Section 4.1
\cite{HuJoSchmiLai}.  On the other hand, the contribution to the
chargino dispersion relations stemming from the variation of the
complex phases in ${\bf M}_D$ only requires explicit CP violating
phases in $\mu$ or $M_2$. Eq.~(\ref{fwkb15}) demonstrates that even
though the phases in the two terms entering $[M_D]_{11,22}$ are
position independent, their contribution to the resulting phase varies
due to the change in the (real) Higgs vevs.
%
%
%
%
%
%
%
%
%
%
\subsection{Diffusion equations and the baryon asymmetry}
In this Section we study the coupled differential equations that
describe particle interactions and transport during the phase
transition. We treat the plasma as consisting of quasi-classical
particles with definite canonical position and momentum.  The phase
space distributions $f_i(\vec{x},\vec{p},t)$ of the particles evolve
according to the classical Boltzmann equation
\begin{equation} \label{diffeq1}
d_tf_i=(\partial_t+\dot{\vec{x}}\cdot\partial_{\vec{x}}+
\dot{\vec{p}}\cdot\partial_{\vec{p}})f_i={\cal C}_i[f].
\end{equation}
The time derivatives of position and momentum obey the Hamilton
equations $\dot{\vec{x}}=\partial_{\vec{p}}E(\vec{x},\vec{p})$ and
$\dot{\vec{p}}=-\partial_{\vec{x}}E(\vec{x},\vec{p})$, where
$E(\vec{x},\vec{p})$ are the dispersion relations derived in the
previous section.  This treatment is an approximation to quantum
Boltzmann equations which have to be discussed in principle.  This
picture is justified for thick walls ($p \gg 1/L_w$) if it predicts a
sizable effect, not dominated by non-leading terms in the derivative
expansion \cite{QBE}.

The Boltzmann equation can in principle be solved
numerically. However, to make it analytically tractable we use the
fluid-type truncation \cite{JPT1}
\begin{equation} \label{diffeq3}
f_i(\vec{x},\vec{p},t)=\frac{1}{e^{\beta(E_i-v_ip_z-\mu_i)}\pm 1}
\end{equation}
for the phase space densities of fermions (+) and bosons (--) in the
rest frame of the plasma.  Here $v_i$ and $\mu_i$ denote the velocity
perturbations and chemical potentials for each fluid, respectively.
We also split $E_i$ into a dominant part $E_{0i}=\sqrt{p^2+m_i^2}$ and
a perturbation $\Delta E_i\sim\partial_z\theta$ which is related to
CP violation.  The chemical potentials are the central quantities that
finally will determine the baryon asymmetry. The velocity perturbation
on the other hand, is only introduced to allow the particles to move
in response to the force, giving rise to chemical potential
perturbations.

We are looking for a ``stationary'' solution of the Boltzmann
equation, because at late times the wall moves with constant velocity
$v_w$. This means that any explicit time dependence enters in the
combination $\bar z\equiv z-v_wt$.  Inserting the fluid ansatz into
the Boltzmann equation (\ref{diffeq1}), eliminating $v_i$ and taking
the difference between particles and anti-particles, we obtain to
linear order in the perturbations $\Delta E_i$ and $\mu_i$, and to
leading order in the wall velocity \cite{CJK}
\begin{eqnarray} \label{diffeq12}
-\kappa_i(D_i\mu_i''+v_w\mu_i')+\sum_p\Gamma_p^d\sum_j\mu_j=S_i,
\nonumber\\
S_i=\frac{D_iv_w}{\langle p_z^2/E_0 \rangle_0}
\langle p_z\Delta E_i'\rangle'
\end{eqnarray}
where primes denote $\partial_{\bar z}$. The diffusion constants
read $D_i=\kappa_i\langle p_z^2/E_0\rangle_0^2/(\bar p_z^2\Gamma_i^e)$,
and $\Gamma^e$, $\Gamma^d$ are the
rates for elastic and inelastic processes, respectively.
The averages are carried out according to
\begin{equation}  \label{diffeq7}
\langle \cdot \rangle \equiv \frac{\int d^3pf_{\pm}'(\cdot)}{\int d^3pf'_+(m=0)}
\equiv \kappa_i\frac{\int d^3pf_{\pm}'(\cdot)}{\int d^3pf'_{\pm}}.
\end{equation}
Here $f_{\pm}'=df_{\pm}/dE_0=-\beta e^{\beta E_0}/(e^{\beta E_0}\pm
1)^2$ is the derivative of the unperturbed Fermi-Dirac or
Bose-Einstein distribution.  The statistical factor $\kappa$ is 1 for
massless fermions, 2 for massless bosons and exponentially small for
particles much heavier than $T$. The subscript ``0'' denotes averaging
with the massless, unperturbed Fermi-Dirac distribution.

The CP violating source term $S_i$ in (\ref{diffeq12}), which is due
to the semi-classical force, is proportional to the diffusion
constant. This is because particles must move in order to build up
perturbations.  In $S_i$ the thermal averages over the CP violating
energy perturbations $\Delta E$ are performed using the massive
distribution functions in order to account for Boltzmann suppression
of heavy particles.

We obtain the transport equations of the MSSM from
Eq.~(\ref{diffeq12}) by specifying the relevant particle species and
interactions in the hot plasma. The network of equations can be
simplified considerably by using conservation laws and neglecting
interactions that are slow compared to the relevant time scale, which
requires $D/v_w^2\ll\Gamma^{-1}$.

In a first step we neglect the weak sphaleron interaction with rate
$\Gamma_{ws}$, which will be included at the end of the calculation.
In the following we will therefore assume baryon and lepton number
conservation.  The neglect of the weak sphalerons allows us to
completely forget about leptons in our transport equations and compute
only the quark and Higgs densities.

We assume the supergauge
interactions to be in equilibrium. The chemical potential of
any particle is then equal to that of its superpartner, and
it is convenient to define the chemical potentials
$\mu_{U}=(\mu_{u^c}+\mu_{\tilde u^c})/2$,
$ \mu_{Q_1}=(\mu_u+\mu_d+\mu_{\tilde u}+\mu_{\tilde d})/4$,
$\mu_{H_1}=(\mu_{H_1^0}+\mu_{H_1^-}+\mu_{\tilde h_1^0}+
\mu_{\tilde h_1^-})/4$, etc.
Furthermore, we take into account the following interactions
 \begin{eqnarray}
&&(\Gamma_y+\Gamma_{yA})(\mu_{H_2}+\mu_{Q_3}+\mu_T), \quad
\Gamma_{y\mu}(\mu_{H_1}-\mu_{Q_3}-\mu_T), \quad
\nonumber\\
&&\Gamma_{ss}(2\mu_{Q_3}+2\mu_{Q_2}+2\mu_{Q_1}+
\mu_T+\mu_B+\mu_C+\mu_S+\mu_U+\mu_D), ~~~~~~~~~~~~
\nonumber\\
 \label{diffeq19}
&&\Gamma_{hf}(\mu_{H_1}+\mu_{H_2}), \quad
\Gamma_m(\mu_{Q_3}+\mu_T), \quad \Gamma_{H_1}\mu_{H_1}, \quad
\Gamma_{H_2}\mu_{H_2}.~~~~~~~~~~~~~~
\end{eqnarray}
The rates in the first line are related to the Lagrangian
\begin{eqnarray}\label{diffeq17}
{\cal L_{\rm int}}&=&y_t t^cq_3H_2+y_t \tilde t^cq_3\tilde h_2 +
y_t t^c\tilde q_3\tilde h_2 - y_t\mu \tilde t^{c*}\tilde q_3^*H_1
+y_tA_t\tilde t^c\tilde q_3H_2+\mbox{h.c.} \qquad
\end{eqnarray}
$\Gamma_{ss}$ denotes the strong sphaleron rate. $\Gamma_{hf}$ is due
to Higgsino helicity flips induced by the $\mu \tilde h_1\tilde h_2$
term. $\Gamma_{H_{1,2}}$ and $\Gamma_m$ correspond to Higgs and axial
top number violating processes, present only in the phase boundary and
the broken phase.

If the system is near thermal equilibrium, number densities and
chemical potentials are related by
\begin{equation}\label{diffeq21}
n_i=\frac{1}{6}k_i\mu_iT^2
\end{equation}
where $k_i$ is the appropriate sum over statistical factors
$\kappa$ introduced in (\ref{diffeq7}),
e.g.~
$k_{Q_1}=N_c(\kappa_u+\kappa_d+\kappa_{\tilde u}+\kappa_{\tilde d})$,
$k_U= N_c(\kappa_{u^c}+\kappa_{\tilde u^c})$,
$k_{H_1}=(\kappa_{H_1^0}+\kappa_{H_1^-}+\kappa_{\tilde h_1^0}+
\kappa_{\tilde h_1^-})$, etc.
$N_c=3$ denotes the number of colors. In the massless limit
one obtains $k_{Q_{1,2,3}}=18$,  $k_U=k_D=...=k_T=9$, $k_{H_{1,2}}=6$.

Using baryon number conservation and neglecting the small Yukawa
couplings of the first and second family quarks the strong sphaleron
rate reads
\begin{eqnarray} \label{diffeq24}
\Gamma_{ss}(2\mu_{Q_3}+\dots+\mu_D)=
\Gamma_{ss}\left[\left(2+9\frac{k_{Q_3}}{k_B}\right)\mu_{Q_3}+
\left(1-9\frac{k_T}{k_B}\right)\mu_T\right]
\end{eqnarray}
To arrive at this expression we made the assumption that all
the squark partners of the light quarks are degenerate in mass.
Assuming equilibrium for the strong sphalerons we obtain
\begin{equation}\label{mssm1}
\mu_T=\frac{2k_B+9k_{Q_3}}{9k_T-k_B}\mu_{Q_3}.
\end{equation}
The reduced set of diffusion equations for the relevant particle species
then finally reads \cite{HS00}
\begin{eqnarray}\label{diffeq25}
-A{\cal D}_{q}\mu_{Q_3}+(\Gamma_y+\Gamma_{yA})[\mu_{H_2}+B\mu_{Q_3}]-
\Gamma_{y\mu}[\mu_{H_1}-B\mu_{Q_3}]+B\Gamma_m\mu_{Q_3}=0~~~~~~
\nonumber \\[.1cm]
-k_{H_1}{\cal D}_h\mu_{H_1}+\Gamma_{y\mu}[\mu_{H_1}-B\mu_{Q_3}]
+\Gamma_{hf}(\mu_{H_1}+\mu_{H_2})+\Gamma_{H_1}\mu_{H_1}=S_{H_1}~~~
\nonumber \\[.1cm]
-k_{H_2}{\cal D}_h\mu_{H_2}+(\Gamma_y+\Gamma_{yA})[\mu_{H_2}+B\mu_{Q_3}]
+\Gamma_{hf}(\mu_{H_1}+\mu_{H_2})+\Gamma_{H_2}\mu_{H_2}=S_{H_2}~~~
\end{eqnarray}
where
\begin{eqnarray}\label{mssm2}
A&=&\frac{9k_Tk_{Q_3}+9k_Bk_{Q_3}+4k_Bk_T}{9k_T-k_B}
\nonumber \\
B&=&\frac{k_B+9k_T+9k_{Q_3}}{9k_T-k_B}
\end{eqnarray}
and ${\cal D}_i\equiv D_i\frac{d^2}{d\bar z^2}+v_w\frac{d}{d\bar z}$.

We keep the rates related to the top Yukawa interactions finite.  If
these interaction are in equilibrium, the resulting diffusion
equations are sourced only by the combination $S_{H_1}-S_{H_2}$,
because of the constraint $\mu_{H_1}+\mu_{H_2}=0$.  As a result the
dominant contribution to the chargino source terms stemming from the
$\theta$-dependent part in the dispersion relations (\ref{fwkb13})
cancels, because the corresponding terms for $\tilde h_1^-$ and
$\tilde h_2^+$ are exactly of the same size.  This is not true for
the $\gamma',\delta'$ contributions which arise from the ``flavor''
transformations ${\bf U}$ and ${\bf V}$. However,  they are suppressed
by the small variation of $\tan\!\beta$ in the bubble wall
\cite{John3,SecoNum,Cline12}.

The full diffusion equations (\ref{diffeq25}) have already been
studied in ref.~\cite{CJKneu}, with source terms corresponding to the
helicity part of the dispersion relations (\ref{fwkb13}). In the
following we also include the flavor part of the source term in the
analysis.

Before solving the network of diffusion equations, we turn to baryon
number generation by the weak sphaleron processes. The evolution of
the baryon number density $n_{\cal B}$ is governed by
\begin{equation} \label{bar3}
-{\cal D}_qn_{\cal B}+3\Theta(\bar z)\Gamma_{ws}(T^2\mu_{{\cal B}_L}-an_{\cal B})=0,
\end{equation}
where we have assumed identical diffusion constants for all quarks and
squarks, and neglected contributions of leptons.  The position
dependence of the weak sphaleron rate is modeled by a step function
$\Theta (\bar z)$: anomalous baryon number violation is unsuppressed
in the symmetric phase $(\bar z>0)$ and suddenly switched off in the
broken phase $(\bar z<0)$.  Baryon number generation is sourced by the
chemical potential of left-handed quarks
\begin{equation}\label{mubl}
\mu_{{\cal B}_L}=C\mu_{Q_3}\equiv
\left[1-\frac{k_{Q_3}+2k_T}{9k_T-k_B}\left(\frac{2k_B}{k_{Q_1}}+
\frac{2k_B}{k_{Q_2}}\right)\right] \mu_{Q_3}.
\end{equation}
The second term in Eq.~(\ref{bar3}) describes damping of the baryon
asymmetry by weak sphalerons in the symmetric phase. The parameter $a$
depends on the degrees of freedom present in the hot plasma.  Taking
only the right-handed stop to be light gives $a=48/7$ \cite{CJKneu}.

From Eq.~(\ref{bar3}) one can easily obtain the baryon to entropy ratio
in the broken phase
\begin{equation}   \label{bar5}
\eta_B\equiv \frac{n_{\cal B}}{s}=\frac{135 \Gamma_{ws}}{2\pi^2g_*v_wT}
\int_0^{\infty}d\bar z \mu_{{\cal B}_L}(\bar z)e^{-\nu\bar z}
\end{equation}
where we have taken the entropy density $s=(2\pi^2g_*/45)T^3$ and
$\nu=3a\Gamma_{ws}/(2v_w)$ \cite{CJKneu}.
$g_*\sim 126$ is the effective number of degrees of freedom at the
phase transition temperature. Eq.~(\ref{bar5}) shows that the
integral over the left-handed quark number, $n_L\propto\mu_{{\cal B}_L}$
in the symmetric phase determines the final baryon asymmetry.

We now return to eqs.~(\ref{diffeq25}) in order to compute $\mu_{{\cal
B}_L}$. These linear second order differential equations can be solved
by finding the appropriate Greens function. In contrast to
ref.~\cite{CJKneu} we do not make any approximations in computing this
Greens function.  We keep the discussion general and consider the
following set of $N$ coupled diffusion equations
\begin{equation}  \label{bar6}
\left(\begin{array}{ccc}
-k_{11}{\cal D}_{11} +\Gamma_{11} & \cdots & -k_{1N}{\cal D}_{1N} +\Gamma_{1N} \\ 
\vdots & \ddots & \vdots \\
-k_{N1}{\cal D}_{N1} +\Gamma_{N1} & \cdots & -k_{NN}{\cal D}_{NN} +\Gamma_{NN}
\end{array}\right)
\left(\begin{array}{c} \mu_1 \\ \vdots \\ \mu_N \end{array}\right)=
\left(\begin{array}{c} S_1 \\ \vdots \\ S_N \end{array}\right)
\end{equation}
where ${\cal D}_{ab}=D_{ab}\frac{d^2}{d\bar z^2}+v_w\frac{d}{d\bar z}$.
The corresponding boundary conditions
read $\mu_a(|\bar z|\rightarrow\infty)=0$.
The matrix valued Greens  function $G_{ab}$  is defined by
$\sum_{c=1}^N(-k_{ac}{\cal D}_{ac} +\Gamma_{ac})G_{cb}(\bar z)=
\delta_{ab}\delta(\bar z)$.
In the transport equations (\ref{diffeq25})
position dependent rates are present, e.g.~$\Gamma_m$.
They typically vanish in the symmetric phase and become maximal in the
broken phase. In order to keep the problem analytically tractable
we  simply model the position dependence of these rates by
step functions, i.e.~$\Gamma_{ab}(\bar z)=\Gamma_{+ab}\Theta(\bar z)+
\Gamma_{-ab}\Theta(-\bar z)$.
%
%
%
%
\section{Numerical results}
\label{sec:final}
In this section we present our numerical results for the
baryon asymmetry in the MSSM, which has to be compared
with the observational value $2-7\times 10^{-11}$ \cite{eta}.
We model the Higgs field $H_1(z)$
in the bubble wall by a kink with width $L_w$ and
take $H_2(z)=\tan\!\beta(z)H_1(z)$, where $\tan\!\beta$
varies in the wall. In all our evaluations we assume $v_c=120$ GeV
and $\tan\!\beta(T_c)=3$. For the critical temperature we
take $T_c=90$ GeV.

In Section 5.3 we emphasized the impact of the squark spectrum
on the baryon asymmetry. A strong phase transition requires
the right-handed stop to be light. In the transport equations we
simply treat it as a massless particle, i.e.~$k_T=18$. We assume
all other sfermions to be heavy compared to $T_c$, i.e.~$k_{Q_3}=k_B=9$.
From Eq.~(\ref{mssm2}) we find for the effective statistical factors
in the transport equation $A=126/13$ and $B=23/13$, and
from Eq.~(\ref{mubl}) $C=5/13$.  This non-universal
squark spectrum has the additional virtue of relaxing the strong
sphaleron suppression of the baryon asymmetry \cite{giush}.
The Higgs fields are treated as massless as well, i.e~$k_H=6$.

For the diffusion constants of quarks and Higgses we take
$D_h=110/T$ and $D_q=6/T$. The rates $\Gamma_{yA}$ and
$\Gamma_{y\mu}$ which involve heavy squarks are
Boltzmann suppressed and we set them to zero in the
following \cite{seco00}. Furthermore we take $\Gamma_y=0.015T$,
$\Gamma_{hf}=0.016T$,  $\Gamma_{m}=0.05T\theta(-z)$,
$\Gamma_{H_1}=\Gamma_{H_2}=0.05T\theta(-z)$. The
weak sphaleron rate is $\Gamma_{ws}=20\alpha_w^5T$.

In the following we separately present the $H_1-H_2$ and  $H_1+H_2$
contributions to the baryon asymmetry, which originate from
the flavor and helicity contributions to the chargino dispersion relation,
respectively. We use the dispersion relations in the canonic momentum.
In the formulation with the kinetic momentum the $H_1-H_2$
part vanishes \cite{CJKneu}. However, a $H_1-H_2$ contribution
has been found in real time Green's function treatments of the
chargino current, most recently in ref.~\cite{seco00}. It would
we very interesting to understand more clearly the relation between our
$H_1-H_2$ source and the one found in that
approach.

\begin{figure}[t]
\begin{picture}(200,140)
\put(0,-55){\epsfysize=5cm{\epsffile{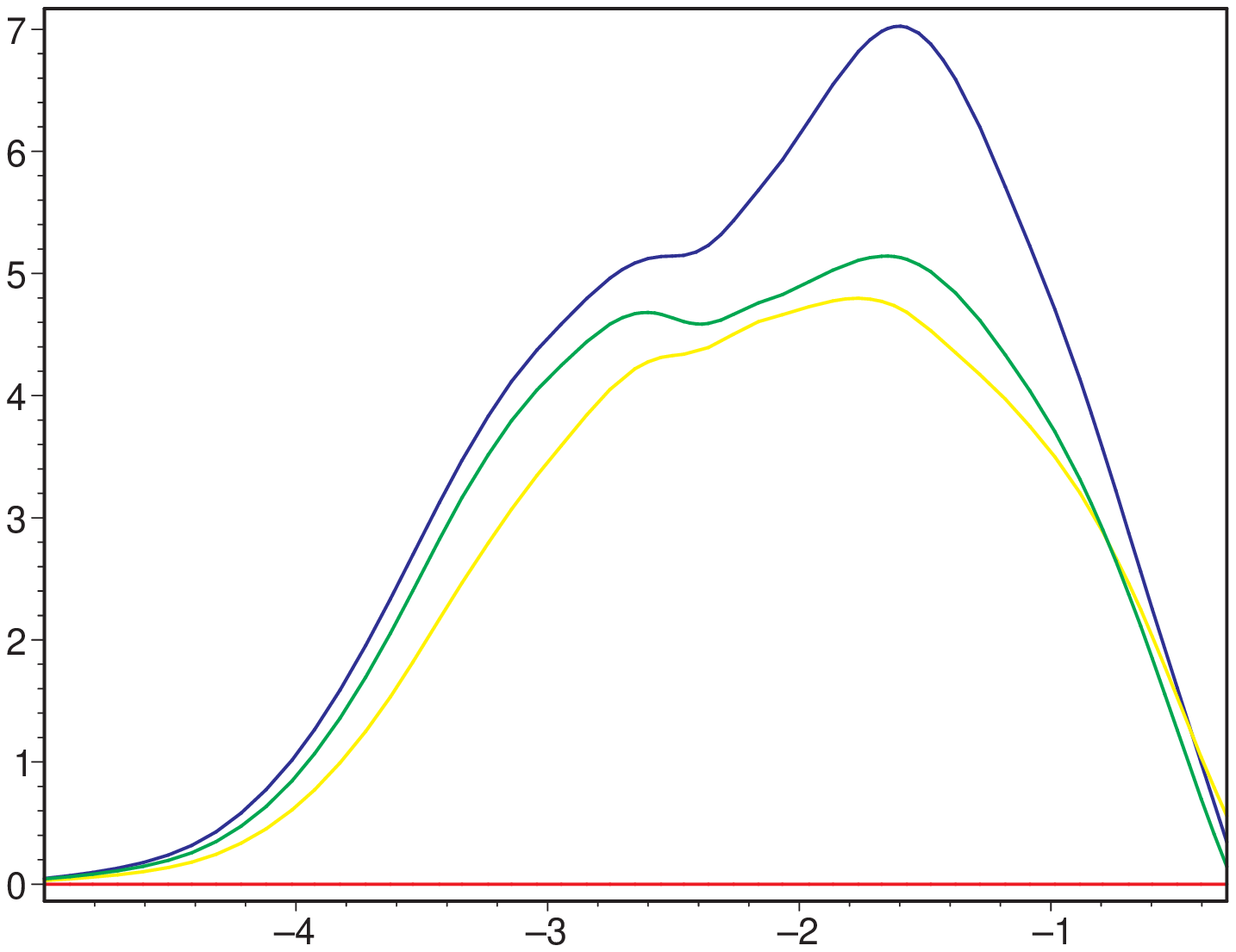}}}
\put(220,-55){\epsfysize=5cm{\epsffile{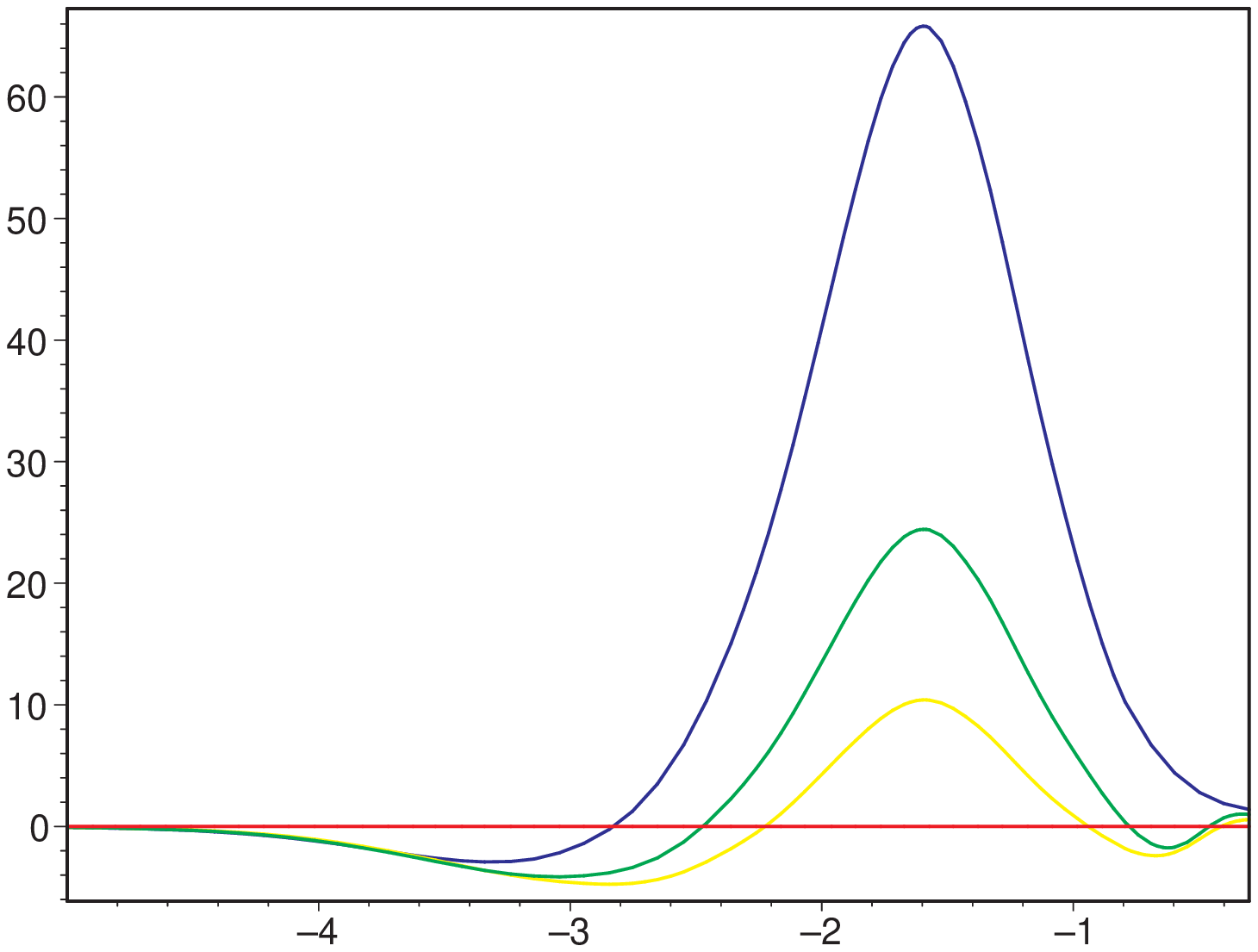}}}
\put(110,-5){$\ln_{10}(v_w)\rightarrow $}
\put(330,-5){$\ln_{10}(v_w)\rightarrow $}
\put(-370,180){$\frac{\eta_B}{2\times10^{10}}\uparrow$  }
\put(15,130){{$H_1+H_2$}}
\put(240,130){{$H_1-H_2$}}
\put(200,100){{$\eta \uparrow$}}
\end{picture}
\caption{
$H_1\pm H2$  contributions to the baryon asymmetry dependent
on $v_w$ for different values of the wall thickness $L_w=20/T,15/T,10/T$
(from below) and $|\mu|=|M_2|=150$ GeV,
arg$(\mu M_2)=\pi/2$ and $\delta\beta=0.01$. $\eta$ is given in units
of $2\times 10^{-11}$ (observational bound).
}
\label{f_eta}
\end{figure}

In fig.~\ref{f_eta} we summarize our results for the baryon
asymmetry generated during the phase transition. To maximize
the result we take $|\mu|=|M_2|$ and maximal CP violation
arg$(M_2\mu)=\pi/2$. For smaller phases the result simply
scales with $\sin({\rm arg}(M_2\mu))$. Moreover, the $H_1-H_2$
result is proportional to the change in the Higgs vev ratio in the
bubble wall, which we take to be $\delta\beta=0.01$.
We work with $\mu=150$ and $\tan\!\beta=3$. The mass of the
lightest chargino eigenstate is then about 110 GeV, in
agreement with the experimental constraints.

In our evaluations we vary the wall velocity in the interval
$10^{-5}<v_w<0.5$. For small values of $v_w$ we deal with a quasi
equilibrium situation, and the baryon asymmetry goes to zero.  Also
large wall velocities suppress $\eta$, since transport in front of the
bubble wall becomes inefficient. From Fig.~\ref{f_eta} we observe that
baryogenesis is most efficient if $v_w$ is a few times $10^{-2}$. This
behavior was already found in refs.~\cite{CJKneu,seco00}.  Most
interesting, calculations of the wall velocity in the MSSM lead to
$v_w$ indeed in that range \cite{GMoo,JSwall}. The detailed
calculations of ref.~\cite{JSwall} lead to\footnote{Potential changes in
$v_w$ due to reheating \cite{ariel} were neglected.}
$v_w=2-3\cdot 10^{-2}$, very near to the maximum for $\eta$.

We also present results for different wall widths $L_w=10/T,15/T,20/T$.
As the derivatives with respect to the $\bar{z}$ coordinate in
the source term suggest, thinner walls induce a larger
baryon asymmetry. However, this effect is much more pronounced
for the $H_1-H_2$ contribution.

For the $H_1+H_2$ contribution we find a baryon asymmetry
which in the most favorable case is about 7 times larger than
the observed value. Using kinetic
momentum in the averaging, one obtains
maximally (in the massless case) a further
factor 2. As result, the assumption of maximal
CP violation can only mildly be relaxed. Large complex phases
are necessary, which are only compatible with EDM experiments
if accidental cancellations happen or the first and second
generation squarks have TeV scale masses. The $H_1-H_2$ contribution
to $\eta$ is an order of magnitude larger, and smaller phases
of the order $10^{-2}$ would be sufficient.

Let us finally compare our results with those of
refs.~\cite{CJKneu,seco00}, which also started from the transport
equations (\ref{diffeq25}).  The authors of ref.~\cite{CJKneu} also
use the semi-classical method but consider only the $H_1+H_2$
contribution, and take somewhat different values for the interaction
rates and diffusion constants. Our results are qualitatively
similar.\footnote{ taking into account a numerical normalization error
corrected in a forthcoming revised version of \cite{CJKneu}.}  The
dependence of the baryon asymmetry on $v_w$ shows small
deviations. This may be due the fact, that in contrast to
ref.~\cite{CJKneu} we exactly calculate the Greens function associated
with the transport equation (\ref{diffeq25}).

Ref.~\cite{seco00} is based on quantum transport equations.  In this
derivation of the chargino source terms also a $H_1-H_2$ contribution
appears. The results in amplitude and velocity dependence are very
similar to those presented here for the $H_1-H_2$ source.  The small
deviations in the regime of small $v_w$ may again be due to an
approximation in the computation of the Greens function if
ref.~\cite{seco00}. It remains to be seen if the $H_1-H_2$
contributions in the approach of
refs. \cite{seco00,irgendwas_von_Riotto,non-eq} and in the
semi-classical method are related, and if they are indeed artificial
as suggested by the use of kinetic momentum.
%
%
\section{Conclusions}
In the first part we have given in some detail the procedure of
solving numerically for the wall profile in the case of more then one
Higgs field for a given effective potential. This extends a previous
description \cite{HuJoSchmiLai}. Within this method we can also
systematically search for CP violating solutions. It turns out -- and
this is confirmed in ref. \cite{LaiRuCPlat} -- that in the MSSM there
is no indication of a spontaneous CP violation.  However, in the
larger parameter space of the NMSSM \cite{HS00} such solutions can be
found in \cite{HuJoSchmiLai}. In the MSSM we remain with the
possibility of explicit CP violating interactions, in particular of
the charginos.

In the second part we treat transport around the stationary proceeding
wall in the quasi-classical approximation. The solution of the set of
Dirac equations for the charginos in the WKB approximation shows a
split in the dispersion relations between particles and anti-particles
at order $\hbar$ because of CP violation.  We have written the full
dispersion relations according to ref.~\cite{CJK} for canonical
momenta, but also presented the relations for kinetic momenta. In
ref.~\cite{CJKneu} arguments have been presented that one should use
the kinetic momentum in Boltzmann transport equations. This would
destroy all source terms of CP violation antisymmetric in the two
Higgses which have been considered to be most important in older work
on the MSSM.  Also in order to compare with work in the spirit of
quantum Boltzmann equations \cite{irgendwas_von_Riotto,non-eq,seco00}
we have kept open for both versions.

Of course one also has to test the two different types of momentum in
averaging procedures, but this just gives a factor (about 2). Indeed,
because $\tan\!\beta$ only varies very weakly in the wall, the
contribution of the symmetric Higgs combination can not be neglected
anyway. However, if the antisymmetric source is absent, rather large
explicit CP violation is needed to obtain the observed baryon
asymmetry even for favorable parameters of the MSSM. Thus the
discussion of consistency with experimental EDM bounds due to
accidental cancellations becomes relevant. On the other side this
might point to an extension of the MSSM, e.g. to some version with a
singlet. It is encouraging that we at least obtained a baryon
asymmetry which is of the right order of magnitude compared to
observations.
\paragraph{Acknowledgement} 
We would like to thank M.~Laine for collaboration in
\cite{HuJoSchmiLai}, technical support, and useful discussions.  This
work was partly supported by the TMR network {\em Finite Temperature
Phase Transitions in Particle Physics}, EU contract no.\
FMRX-CT97-0122. S.~H.~is supported in part by the Alexander von
Humboldt Foundation. P.~J.~is supported by the Deutsche
Forschungsgemeinschaft.
%
%
%
%

\end{document}